\documentclass[reprint,superscriptaddress,longbibliography,amsmath,amssymb,aps,prx,nofootinbib]{revtex4-1}
\usepackage{graphicx}
\usepackage{dcolumn}
\usepackage{bm}
\usepackage{soul}
\usepackage{afterpage}

\usepackage{float}
\usepackage{placeins}
\usepackage{dsfont}
\usepackage[utf8]{inputenc}
\usepackage{bm}
\usepackage{amssymb}
\usepackage{amsmath}
\usepackage{amsfonts}
\usepackage{graphicx}
\usepackage[usenames,dvipsnames]{xcolor} 
\usepackage{dcolumn}
\usepackage{bm}
\usepackage[pdftex,
            pdftitle={SLIOMs},
            bookmarks,
            colorlinks,
            linkcolor=black,
            citecolor=magenta,
            menucolor=black,
            urlcolor=blue,
            plainpages=false,
            pdfpagelabels,
            hypertexnames=false]{hyperref}

\newcommand{\avg}[1]{\left< #1 \right>}

\newcommand{\ket}[1]{\left| #1 \right>}
\newcommand{\bra}[1]{\left< #1 \right|}
\newcommand{\braket}[1]{ \langle{{#1}}\rangle}

\let\baraccent=\= \renewcommand{\=}[1]{\stackrel{#1}{=}}

\usepackage[normalem]{ulem}

\usepackage{framed}
\definecolor{shadecolor}{gray}{0.95}

\begin{document}
\title{Statistical localization: from strong fragmentation to strong edge modes}

\author{Tibor Rakovszky }
\thanks{These authors contributed equally to this work.\\}
\affiliation{Department of Physics, Technical University of Munich, 85748 Garching, Germany}
\affiliation{Munich Center for Quantum Science and Technology (MCQST), Schellingstr. 4, D-80799 M\"unchen, Germany }
\affiliation{Kavli Institute for Theoretical Physics, University of California, Santa Barbara, CA 93106, USA}
%
\author{Pablo Sala}
\thanks{These authors contributed equally to this work.\\}
\affiliation{Department of Physics, Technical University of Munich, 85748 Garching, Germany}
\affiliation{Munich Center for Quantum Science and Technology (MCQST), Schellingstr. 4, D-80799 M\"unchen, Germany }
%
\author{Ruben Verresen}
\affiliation{Department of Physics, Technical University of Munich, 85748 Garching, Germany}
\affiliation{Max-Planck-Institute for the Physics of Complex Systems, 01187 Dresden, Germany}
\affiliation{Department of Physics, Harvard University, Cambridge, MA 02138, USA}
\author{Michael Knap}
\affiliation{Department of Physics, Technical University of Munich, 85748 Garching, Germany}
\affiliation{Munich Center for Quantum Science and Technology (MCQST), Schellingstr. 4, D-80799 M\"unchen, Germany }
\affiliation{Institute for Advanced Study, Technical University of Munich, 85748 Garching, Germany}

\author{Frank Pollmann}
\affiliation{Department of Physics, Technical University of Munich, 85748 Garching, Germany}
\affiliation{Munich Center for Quantum Science and Technology (MCQST), Schellingstr. 4, D-80799 M\"unchen, Germany }

\date{\today }

\begin{abstract}
Certain disorder-free Hamiltonians can be non-ergodic due to a \emph{strong fragmentation} of the Hilbert space into disconnected sectors. 
Here, we characterize such systems by introducing the notion of `statistically localized integrals of motion' (SLIOM), whose eigenvalues label the connected components of the Hilbert space.
SLIOMs are not spatially localized in the operator sense, but appear localized to sub-extensive regions when their expectation value is taken in typical states with a finite density of particles. 
We illustrate this general concept on several Hamiltonians, both with and without dipole conservation.
Furthermore, we demonstrate that there exist perturbations which destroy these integrals of motion in the bulk of the system, while keeping them on the boundary. 
This results in statistically localized \emph{strong zero modes}, leading to infinitely long-lived edge magnetizations along with a thermalizing bulk, constituting the first example of such strong edge modes in a non-integrable model. 
We also show that in a particular example, these edge modes lead to the appearance of topological string order in a certain subset of highly excited eigenstates.
Some of our suggested models can be realized in Rydberg quantum simulators.
\end{abstract}

\maketitle

\tableofcontents
\section{Introduction}

The internal dynamics of closed quantum-many body systems has been a central topic in condensed matter physics over the last decade, with strong connections to quantum information theory. This has been motivated by experimental advances in preparing and manipulating quantum systems that are isolated from their environments to a high precision. An interesting question to ask is how such systems approach thermal equilibrium under their unitary dynamics~\cite{Gring1318,Hild2014,Brown540,Kaufman794,Yijun2018,Brydges2018,ETHreviewRigol16,GogolinReview,Meinert945}. The \emph{eigenstate thermalization hypothesis} (ETH) has emerged as a sufficient condition for thermalization, and has been subsequently demonstrated to hold in a variety of interacting quantum systems~\cite{Deutsch91,Srednicki94,Rigol2008,Kim2014,ETHreviewRigol16}. 

Due to the seeming generality of ETH, much interest has been generated by mechanisms that violate it and lead to a breakdown of thermalization. Such a breakdown can arise due to the existence of an extensive number of conservation laws. One class of models where this occurs are \emph{integrable systems}, where the conserved quantities arise as integrals of local (or quasi-local) densities~\cite{Rigol2007,Kinoshita2006,Essler16}. Interestingly, it has been shown that strong disorder can also lead to an infinite number of emergent conservation laws without the need for fine-tuning, defining the so-called \emph{many-body localized} (MBL) phase~\cite{Basko06,Nandkishore14,AltmanVosk,Schreiber2015}. The integrals of motion in this case (dubbed \emph{local integrals of motion}, or LIOMs for short) are exponentially localized in space around a specific position~\cite{Huse14,Serbyn13cons,Lioms17,Abanin19}. Consequently, the dynamics in MBL systems preserves memory of the initial state locally.

Several works investigated the possibility of mimicking similarly localized behavior without explicitly breaking translation invariance~\cite{Muller2015,Yao2016,PAPIC2015,Smith01,Smith02,Brenes18,Michailidis18,Refael18,Schulz19}, as well as the possibility of intermediate behavior, such as the existence of a small number of ETH-violating eigenstates within an otherwise generic spectrum of states~\cite{Moudgalya01,Moudgalya02,Iadecola2018,Iadecola2019,Neupert2019,ShiraishiMori,TurnerNatPhys,TurnerPRB,Bernien2017,Choi2018,Motrunich2018,Feldmeier19,Schecter2019, Sanjay19,Serbyn19}. Recently, the authors of the present paper, following earlier work on dipole-conserving random circuits~\cite{Pai18},
identified a novel mechanism for such non-ergodic behavior, dubbed \emph{Hilbert space fragmentation}~\cite{Sala19,Vedika19}.
In this scenario, the space of many-body states in some simple local basis splits into exponentially (in system size) many distinct sectors, which are disconnected from one another\footnote{Recently, a model with similar properties was discussed in Ref.~\onlinecite{Patil19}. Unlike the cases we consider here, the fragmentation there is due to explicit local conservation laws.}.
Especially interesting is the case of \emph{strong fragmentation}, where the size of the largest connected sector is exponentially smaller than the total number of states. 
In the particular example discussed in Ref.~\onlinecite{Sala19}, it was found that this can lead to not only a complete breakdown of ETH, but also to effectively localized behavior in the form of infinitely long-lived autocorrelations, similar to true localization. 
However, establishing a clear connection between such localization and the structure of the Hilbert space remained an open challenge.

While Refs.~\onlinecite{Sala19,Vedika19} provided a general mechanism for Hilbert space fragmentation and uncovered many of the intriguing features resulting from it, understanding the nature of the corresponding integrals of motion was left as an open question. In the present work we uncover these conserved quantities in two illustrative cases, focusing on strongly fragmented Hilbert spaces. We also formulate the general principle behind such conserved quantities and discuss both their similarities and their differences compared to the LIOMs of MBL systems. We first consider a simple example that exhibits strong fragmentation (without conserving dipole moment), where we can illustrate the nature of the integrals of motion in a straightforward manner. Later we return to the dipole-conserving minimal model of Ref.~\onlinecite{Sala19} and identify \emph{all} the conserved quantities that label the components of its strongly fragmented Hilbert space. This is achieved via a non-local mapping to a different model with explicit local constraints. We analytically show that these conservation laws lead to spatial localization and finite autocorrelations in the thermodynamic limit.

A unifying feature of the conserved quantities we uncover is what we name \emph{statistical localization}. These are non-local operators, whose expectation values in typical states pick up contributions primarily from specific spatial regions that are sub-extensive in their size. Unlike the case of LIOMs, this region depends on properties of the quantum state in question; in particular, the models we consider possess a conserved U(1) charge and the localization properties of the new integrals of motion turn out to depend on the overall filling fraction. Moreover, while some of these integrals of motion are effectively localized to finite regions in the dipole-conserving case (much like LIOMs), others are only `partially localized', i.e. they correspond to regions that grow sub-linearly with system size. 

Having identified the new conserved quantities, we show that they give rise to another exciting possibility: \emph{statistically localized strong zero modes} localized at the boundaries of a finite system. These are analogous to the strong boundary zero modes (SZM) discussed in the literature~\cite{Fendley_2012,Fendley_2016,Fendley16,Kemp17,Chetan17,Garrahan19}, but unlike previous instances, they occur in non-integrable systems, co-existing with a completely thermalizing bulk. We explicitly construct such zero modes (which commute exactly with the Hamiltonian even for finite systems), by perturbing the strongly fragmented Hamiltonians in specific ways, destroying the integrals of motion in the bulk, while leaving them intact at the boundaries. The resulting models exhibit similar phenomenology as previously studied cases of SZM, with infinite edge coherence times, as well as \emph{exact} degeneracies throughout the spectrum. Our construction provides an example of \emph{exact} strong zero modes in a non-integrable system, stabilized by the dynamical constraints. We also propose an experimental setup for realizing such models with Rydberg atoms in an optical lattice.

Finally, we discuss how in cases with strong Hilbert space fragmentation, the edge modes can lead to the appearance of highly excited states with non-trivial topological string order. This further reinforces the analogy between strong fragmentation and many-body localization, as the latter can also lead to excited states exhibiting forms of order that are not otherwise allowed at finite temperature~\cite{Huse2013,Chandran2014}.

To summarize, our main results are the following.
\begin{itemize}
    \item We introduce the concept of SLIOMs and illustrate their usefulness for two separate models.
    \item Using this concept, we construct experimentally relevant non-integrable models with exact strong zero modes at their edges.
    \item We construct all the SLIOMs for a 3-site dipole-conserving model, and show explicitly that they lead to localized dynamics.
    \item We show that the same conservation laws protect topological string order in a subset of excited states at finite energy densities.
\end{itemize}

The remainder of the paper is organized as follows. In Sec.~\ref{sec:tJz} we provide a detailed discussion of a simple model that exhibits strong fragmentation. We introduce the model in Sec.~\ref{sec:tJz_def} and then construct the full set of conserved quantities that characterize the connected subspaces, using them to illustrate the concept of SLIOMs, which we define in Sec.~\ref{sec:slim}. We describe the effect of SLIOMs on thermalization in the bulk and at the boundary in Sec.~\ref{sec:tJz_thermal}, constructing a perturbed model with strong zero modes and a thermalizing bulk. In Sec.~\ref{sec:H3} we extend our discussion to the strongly fragmented, dipole-conserving Hamiltonian introduced in Ref.~\onlinecite{Sala19}. We use a non-local mapping to analytically construct the complete set of conserved quantities that describe its fragmentation, and discuss both the similarities and differences compared to the model of Sec.~\ref{sec:tJz}. We discuss how the SLIOMs in this case lead to localized dynamics, and discuss the implications for entanglement growth in Sec,~\ref{sec:types}. We comment on the appearance of string order in excited states in Sec.~\ref{sec:spt} before concluding in Sec.~\ref{sec:outlook}.

\section{Illustrative example of SLIOMs: $t-J_z$ model}\label{sec:tJz}

Here we introduce the main concept of our paper, that of \emph{statistically localized integrals of motion} (SLIOM), which are non-conventional integrals of motion responsible for the lack of thermalization in the systems we consider.
It will be useful to contrast these with the well known case of LIOMs~\cite{Huse14,Serbyn13cons,Lioms17}, which play a similar role in MBL systems. 
Such a LIOM $\tau_n^z$ is localized around some given site $n$ in an operator sense: when written as a sum of 'physical' operators, $\tau_n^z=\sum_i \mathcal{O}^n_i$, the spectral norm~\footnote{The spectral norm of an operator $A$ is induced by the $L^2$-norm and takes the form $\|A\| \equiv \textrm{max}_{x\neq 0}\|Ax\|/\|x\| $.} $\|\mathcal{O}_i^n\|$ of $\mathcal{O}_i^n$ that have support on sites far from $n$ is exponentially suppressed\footnote{One usually chooses a complete set of basis operators, for example direct products (`strings') $\mathcal{S}$ of local Pauli operators in the case of a spin-1/2 chain. One can then write $\tau_n^z = \sum_{\mathcal{S}} c^n_{\mathcal{S}} \mathcal{S}$; the Pauli strings all have unit spectral norm, so the exponential (in the spatial support of $\mathcal{S}$) decay is carried entirely by the coefficients $c^n_{\mathcal{S}}$.}.
The operators we consider are not localized in this sense: they are equal weight superpositions of operators with supports of all sizes, i.e., $\|\mathcal{O}_i^n\| \sim \text{const.}$
However, when the expectation values are taken in `typical states' (to be specified below), these values $\braket{{\mathcal{O}_i^{n}}^{\dagger} \mathcal{O}_i^n}$ only pick up contributions from a region that consists of a vanishingly small fraction of the whole system (and whose precise location and width depend on the state in question): hence the term \emph{statistically} localized. 

\begin{figure}
	\includegraphics[width=1.\columnwidth]{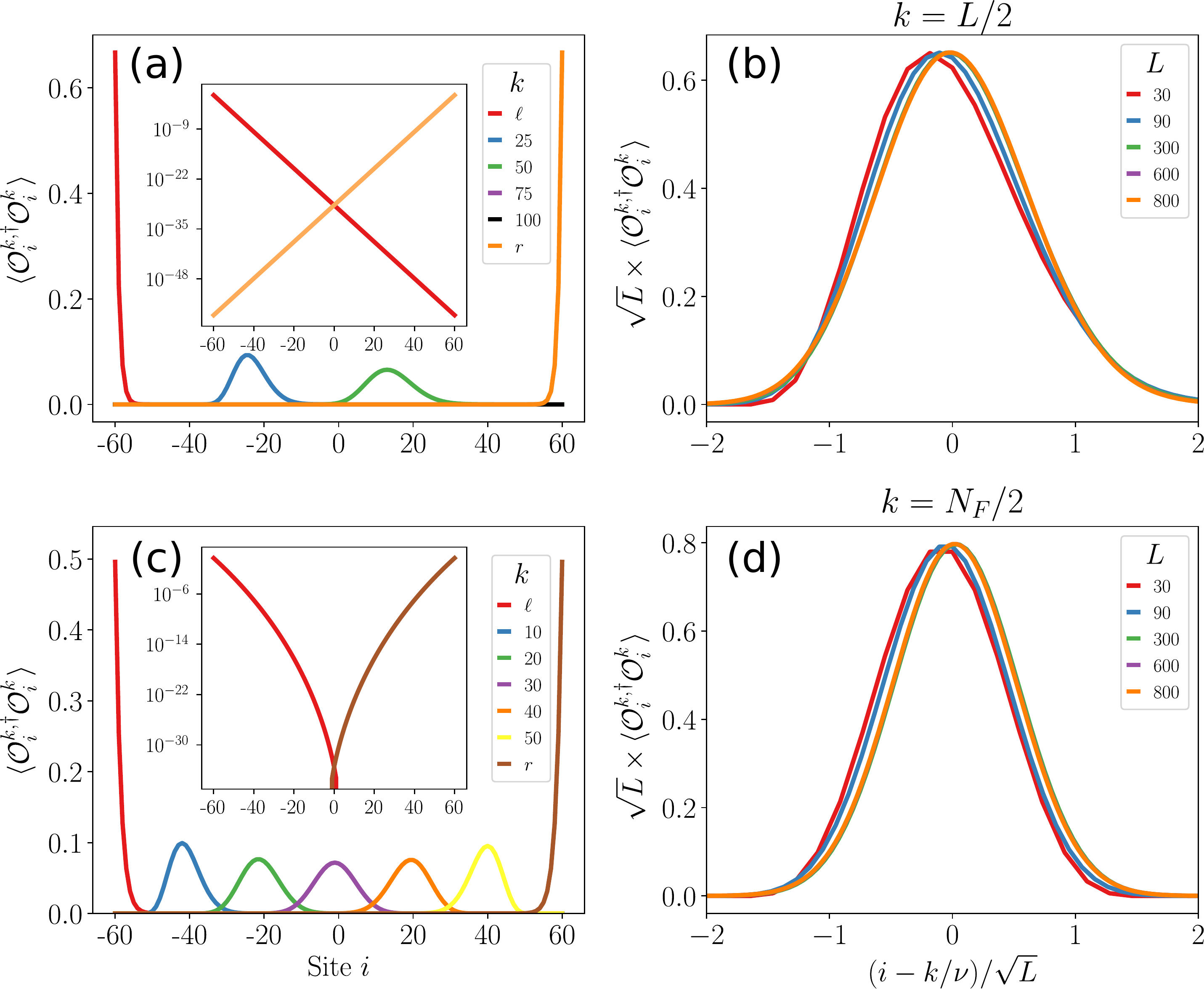}
	\caption{\textbf{Statistical locality of SLIOMs.} 
	Expectation value $\braket{\psi|{\mathcal{O}^{k}_i}^{\dagger}\mathcal{O}^{k}_i|\psi}$ for the string operators appearing in the definition of SLIOMs $\hat q_k = \sum_i \mathcal{O}^{k}_i$, see Eq.~\eqref{eq:qk_def}. The averages are performed over: a Haar random state $\ket{\psi}$ in the full Hilbert space with average filling fraction $\nu=2/3$ in panels (a,b), and a random state with a fixed filling fraction $\nu = N_F/L = 1/2$ in panels (c,d), evaluated analytically via Eqs.~\eqref{eq:Haaravg} and \eqref{eq:U1Haaravg} respectively. (a,c): In both cases, the $k$-th particle, is statistically localized around the average position $\bar i = k/\nu$. \textbf{Inset:} the statistical localization of the boundary SLIOMs $\hat q_{\ell}, \hat q_r$ (defined in Eqs.~\eqref{eq:ql_def} and~\eqref{eq:qr_def}), with at least exponential decay towards the bulk. (b,d): When considering SLIOMs in the bulk, $k \propto L$ ($k=L/2$ for (c) and $k= N_F/2 = L/4$ for (d)), the width of the distribution scales as $\sqrt{L}$, and the height as $1/\sqrt{L}$.}
	\label{fig:fig1}
\end{figure}

\subsection{Definition of the model}\label{sec:tJz_def}

This general concept is best illustrated through a simple example. We consider a one-dimensional Fermi-Hubbard model under the assumption that the Hubbard on-site repulsion is sufficiently strong as to prohibit double occupancy of sites. In this limit, and after replacing Heisenberg by Ising interactions, one obtains the so-called $t-J_z$ model~\cite{Zhang97,Bohrdt2017}. In this work we consider the following simplified version of it\footnote{The definition of the $t-J_z$ model usually includes an additional density-density interaction~\cite{Zhang97}. We drop that term for simplicity, but keeping it would not change the following discussion.}:
\begin{equation}\label{eq:tjz}
H_{t-J_z} = -t \sum_{\substack{i=1 \\ \sigma=\uparrow,\downarrow}}^{L-1} (\tilde{c}_{i,\sigma} \tilde{c}_{i+1,\sigma}^\dagger + \text{H.c.} ) + J_z\sum_{i=1}^{L-1} S^z_i S^z_{i+1},
\end{equation}
where the dressed fermionic operators $\tilde{c}_{i,\sigma}\equiv c_{i,\sigma}(1-c_{i,-\sigma}^\dagger c_{i,-\sigma})$ incorporate the hard-core constraint. $\sigma=\uparrow,\downarrow$ is a spin index, and the on-site constrained Hilbert space consists of only three states: $0,\uparrow,\downarrow$, with $0$ denoting an empty site. 
The first term in Eq.~\eqref{eq:tjz} describes the constrained hopping of fermions and the second term is a nearest neighbor Ising-type interaction with spin operators defined as
\begin{align} \label{eq:SpinOp} 
S_i^{\alpha}=(\tilde{c}_{i,\uparrow}^\dagger,\tilde{c}_{i,\downarrow}^\dagger){\sigma^{\alpha}}(\tilde{c}_{i,\uparrow},\tilde{c}_{i,\downarrow})^T & & \text{with } \alpha=x,y,z;
\end{align}
where we omit a factor of $1/2$ for later convenience.
In our numerics we fix $t=1$ and take $J_z=1/4$, avoiding the integrable point $J_z=0$~\cite{KOTRLA90}. This Hamiltonian conserves both the fermion number, $N_F \equiv \sum_j \left(\tilde{n}_{j,\uparrow} + \tilde{n}_{j,\downarrow}\right)$, and the total spin, $S^z_\text{tot} \equiv \sum_j \left(\tilde{n}_{j,\uparrow} - \tilde{n}_{j,\downarrow}\right)$, with the number operator defined as $\tilde{n}_{j,\sigma} \equiv \tilde{c}_{i,\sigma}^\dagger \tilde{c}_{i,\sigma}$. 

The constrained hopping implies that the dynamics of the model consists entirely of a `re-shuffling' of the hole positions, with the direction of each individual spin always remaining unchanged
~\cite{Batista01,Peres00}.
Thus, for fixed particle number $N_F$, any product state in the $0,\uparrow,\downarrow$ basis is characterized by a pattern of $N_F$ spins, each pointing either up or down.
This pattern is a conserved quantity: only states with the same spin pattern are connected by the dynamics\footnote{A classical, discrete time model with the same symmetries was considered in Refs.~\cite{Medenjak2017,Klobas18}.}.
Therefore, the $3^L$ dimensional many-body Hilbert space fragments into exponentially many disconnected sectors, labeled by the different spin patterns, an example of strong fragmentation~\cite{Sala19,Vedika19}.

In the following we focus on a chain with open boundaries, where the fermions can be labeled by an integer $k$, starting from either the left or the right edge of the system (we discuss periodic boundary conditions in App.~\ref{app:pbc}). 
In this case, the dimension of a given sector is $\binom{L}{L-N_F}=\binom{L}{N_F}$, which counts the number of ways to re-shuffle the $L-N_F$ holes.
Note that the dimension of the largest connected sector, attained for $N_F=L/2$, scales asymptotically as $2^L$ (up to logarithmic corrections), and thus it is a vanishing fraction of the full Hilbert space dimension (as well as of the dimension of the global $(N_F,S^z_\text{tot})$ symmetry sector it is contained in).
For a given $N_F$, there are $2^{N_F}$ different sectors, corresponding to the choices of spin pattern. 
One could easily generalize this model, by allowing for fermions with a larger spin $S$ \cite{Batista00}. 
This would not change the size of the sectors, but increase their number to $(2S+1)^{N_F}$, thus increasing the fragmentation (decreasing the ratio of the largest component to the whole Hilbert space).

Before analyzing the $t-J_z$ Hamiltonian in more detail, let us briefly comment on its relation with various other models. First, we note that while here we focus on a version of the model where no double occupancy is allowed, in fact, the spin pattern is also conserved in the presence of doublons, as long as their total number is conserved due to the strong interactions~\cite{Peres00} (and as long as total spin is conserved as well). Second, we point out that XX spin ladders are known to have subspaces where the dynamics is equivalent to that of $H_{t-J_z}$ with a fixed spin pattern~\cite{Znidaric2013,Iadecola2018}. These can be thought of as \emph{weakly fragmented} analogues of our model, where certain, but not all, spin patterns are conserved. It would be interesting to explore whether the conserved quantities we discuss in the next section have any bearing on the dynamics of these systems. 

\subsection{Statistically localized integrals of motion}\label{sec:slim}

Fixing the complete spin pattern is analogous to fixing the eigenvalues of \emph{all} LIOMs in a many-body localized system, which determines a single eigenstate of the localized Hamiltonian~\cite{Huse14,Serbyn13cons}.
The difference is that the spin pattern only fixes a finite dimensional symmetry subspace, rather than a single many-body state, due to the fact that the holes are free to move.
Therefore the analogue of a \emph{single} LIOM is the operator which measures \emph{the spin of the $k$-th fermion}.
This is our first example of a statistically localized integral of motion, as we now argue.

\textbf{Definition (SLIOM).} By a statistically localized integral of motion (SLIOM) we mean an operator $\hat{q} \equiv \sum_{i=1}^{L} \mathcal{O}_i$ satisfying the following two properties:
\begin{enumerate}
\item $\hat q$ is conserved, $[\hat H, \hat q] = 0$
\item For almost all states $\ket{\psi}$, the expectation value $\braket{\psi|\mathcal{O}_i^\dagger \mathcal{O}_i|\psi} = \|\mathcal{O}_i\ket{\psi}\|^2$, when treated as a probability distribution\footnote{As we shall see below, in the cases we consider $\mathcal O_i^\dagger \mathcal O_i$ is a projector, such that this interpretation is natural. In general, one might need to normalize the distribution to sum up to 1. We ignore the trivial cases when all $\langle \mathcal O_i^\dagger \mathcal O_i \rangle = 0$.} over sites $i$, is localized to a sub-extensive region in space,
\begin{equation}
\frac{\text{Var}(i)}{L} \equiv \frac{\sqrt{\sum_i \braket{\mathcal{O}_i^\dagger \mathcal{O}_i} \, i^2 - \left(\sum_i\braket{\mathcal{O}_i^\dagger \mathcal{O}_i} \, i\right)^2}}{L} \overset{L\to\infty}{\longrightarrow} 0.
\end{equation}
\end{enumerate}
For example, the average global magnetization in a spin-$1/2$ chain, $\frac{1}{L}\sum_i \sigma_i^z$, is not a SLIOM since it has $\text{Var}(i) / L = 1/\sqrt{12}$. In App.~\ref{app:slim_def} we give a slightly different and more refined version of the definition, which captures more of the structure of the conserved quantities we discuss in the following (see also Sec.~\ref{sec:H3labels}).

Some comments are in order. \textbf{i)} In the definition \emph{almost all} is meant in the sense that states $\ket{\psi}$ violating this condition are of measure zero in the thermodynamic limit. 
\textbf{ii)} In the definition we did not specify the form of the operators $\mathcal{O}_i$, except that there is one for each site in the chain and that their sum gives a conserved quantity. 
In the examples below they will turn out to be \emph{string-like} objects, extending from one end of an open chain up to site $i$. 
\textbf{iii)} In the definition, we have characterized localization in a rather weak sense: instead of requiring that the distribution is localized to a finite region, we only required that its width is sub-extensive. 
In the following we will distinguish two cases: the \emph{fully localized} one, where $\text{Var}(i) \sim O(1)$ (which is most similar to MBL) and the \emph{partially localized} one, where $\text{Var}(i) \sim L^\kappa$ for some $0 < \kappa < 1$. 
In fact, we will see that for the $t-J_z$ model, the SLIOMs that are relevant for the bulk are all partially localized with $\kappa = 1/2$. 
This localization is therefore much weaker than the case of MBL, but still has non-trivial consequences for the dynamics, as we will show in Sec.~\ref{sec:tJz_thermal}. 
On the other hand, a subset of the conserved quantities, are in fact localized near the boundaries, and behave very similarly to so-called strong boundary zero modes. 
The dipole-conserving Hamiltonian considered in Sec.~\ref{sec:H3}, however, has fully localized SLIOMs also in the bulk (along with partially localized ones).

\paragraph*{\textbf{Example -- spin configurations in the $t-J_z$ model.}} We now illustrate how the above definition applies to the $t-J_z$ Hamiltonian introduced in Sec.~\ref{sec:tJz_def}. Taking open boundary conditions (OBC), we can define an operator that measures the spin of the $k$-th fermion from the left edge of the chain:
\begin{equation}\label{eq:qk_def}
\hat q_k  \equiv \sum_{i=1}^L \mathcal{O}^k_i = \sum_{i=1}^{L} \hat{\mathcal{P}}^k_i S^z_i,
\end{equation}
where $\hat{\mathcal{P}}^k_i$ is a projection operator, diagonal in the computational basis, that projects onto configurations where the $k$-th charge is exactly on site $i$. The operators $\hat q_k$ form a set of extensively many conserved quantities for $H_{t-J_z}$ with OBC, whose combined eigenvalues label all the different possible spin patterns, such that $\sum_k \hat q_k = S^z_{\text{tot}}$. Each $\hat{q}_k$ has three eigenvalues, $\gamma_k = +1,-1,0$, the latter corresponding to configurations with $ k>N_F$ (consequently, $\hat{q}_k^2$ is a projection onto configurations with $k \leq N_F$). However, not all possible combinations are allowed: if $\gamma_k = 0$ for some $k$ then $\gamma_{k'>k} = 0$ as well. The total number of possible configurations is therefore $\sum_{N_F=0}^{L} 2^{N_F} = 2^{L+1}-1$, each corresponding to one of the connected sectors in the theory. Note that the definition of $\hat q_k$ explicitly breaks spatial parity. One could alternatively define a set of operators starting from the right edge; these encode the same information regarding the block structure of the Hamiltonian. 

As we now argue, the operator $\hat{q}_k$ falls under the above notion of a statistically localized integral of motion, with the role of $\mathcal{O}_i$ in the definition played by the operator $\hat{\mathcal{P}}^k_i S^z_i$. The reason for the statistical localization in this case can be seen intuitively: for a typical state with some average filling $\nu = \avg{N_F}/L$, the $k$-th charge is most likely to be found in the vicinity of position $i = k/\nu$. The width of the distribution should also depend on $\nu$, going to zero in the limit $\nu\to 1$. On the other hand, one can always find atypical states with the same filling where the $k$-th charge is localized at some different position, or not localized at all. To better understand the nature of the conserved quantities $\hat{q}_k$, we now consider their expectation values for two different ensembles of randomly chosen pure states (in App.~\ref{app:eigstates} we also consider specific eigenstates of $H_{t-J_z}$). 

\emph{Global Haar random states --} Let us first consider the case when $\ket{\psi}$ is chosen Haar randomly from the entire Hilbert space~\cite{Reimann07,Steinigeweg15}. This is a state with a fermion density $\nu = 2/3$ on average. We are interested in the average and variance of the expectation value of the operator ${\mathcal{O}^k_i}^\dagger \mathcal{O}^k_i = \hat{\mathcal{P}}^k_i$, which is a projector onto configurations where site $i$ is occupied and the leftmost $i-1$ sites host a total of $k-1$ fermions. When averaged over the Haar ensemble, the expectation value is the same as in an infinite temperature ensemble, simply given by the relative number of such configurations
\begin{equation}\label{eq:Haaravg}
p_\text{Haar}(i;k)\equiv  \mathbb{E}_\text{Haar}[\braket{\psi|{\mathcal{O}^k_i}^\dagger \mathcal{O}^k_i|\psi}]  = \nu^{k}(1-\nu)^{i-k} \binom{i-1}{k-1},
\end{equation}
for $i\geq k$ and $\nu=2/3$. $\sum_i p_\text{Haar}(i;k)$ is the probability of having at least $k$ charges in the system; we focus on $k / L < \nu$, in which case this probability is exponentially close to $1$. 

The distribution $p_\text{Haar}$ is peaked around the position $\bar i = k/\nu$. For the leftmost charge ($k=1$), it simply decays exponentially into the bulk as $\sim 3^{-i}$. In general, for a fixed finite value of $k$, $p_\text{Haar}(i;k)$ is independent of the system size $L$ and has some finite width. However, to probe the bulk of the system, one should choose $k = \alpha L$ for some constant $0 < \alpha < \nu$. In this case, due to the binomial coefficient, the distribution has a standard deviation that scales with system size as $\sim \sqrt{L}$. Nevertheless, it is still `partially localized' in the sense defined previously, such that the width relative to the system size vanishes as $1/\sqrt{L}$ in the thermodynamic limit. This is shown in Figs.~\ref{fig:fig1}(a-b). Outside of the $O(\sqrt{L})$ region, the distribution has a tail that falls off asymptotically faster than exponentially. To leading order in the thermodynamic limit, $L\to\infty$ and for $x\equiv i/L \geq \alpha$, the distribution becomes $ \propto \exp\left[-L\left(x\log{3} -\alpha\log{2} - x h_2(\alpha/x)\right)\right]$, where $h_2(\lambda) \equiv -\lambda\log{\lambda} - (1-\lambda)\log(1-\lambda)$ is the binary entropy function. Note that the exponent vanishes when $x=\alpha/\nu=3\alpha/2$ and is negative otherwise. 

Similarly, one can calculate the variance over choices of Haar random states (see App.~\ref{app:Haar} for details). This gives
\begin{multline}\label{eq:HaarVar}
\mathbb{E}_\text{Haar} [|\braket{\psi|{\mathcal{O}^k_i}^\dagger \mathcal{O}^k_i|\psi}|^2] - |\mathbb{E}_\text{Haar}[\braket{\psi|{\mathcal{O}^k_i}^\dagger \mathcal{O}^k_i|\psi}]|^2 = \\ 
= \frac{1}{3^L + 1} \left[p_\text{Haar}(i;k) - p_\text{Haar}(i;k)^2\right],
\end{multline}
which is exponentially suppressed compared to the average, indicating that indeed the vast majority of states in the Hilbert space gives rise to very similar distributions for $\braket{{{\mathcal{O}^k_i}^\dagger \mathcal{O}^k_i}}$.

\emph{Random states with fixed particle number --} While the above calculation shows that most states lead to a sharply peaked distribution, it is also natural to consider states that are randomly chosen within a sector with fixed total fermion number $N_F$. As we now show, the distributions in this case are still (partially) localized in space, but their location and width now depends explicitly on the filling fraction $\nu = N_F / L$, emphasizing the statistical nature of the localization. One can perform the averaging over the restricted Haar ensemble (see App.~\ref{app:Haar}) to obtain
\begin{equation}\label{eq:U1Haaravg}
 p_{N_F}(i;k) \equiv \mathbb{E}_{N_F}[\braket{\psi|{\mathcal{O}^k_i}^\dagger \mathcal{O}^k_i|\psi}] = \frac{\binom{i-1}{k-1}\binom{L-i}{N_F-k}}{\binom{L}{N_F}}.
\end{equation}
This distribution differs from the previous one in several aspects. First, $p_{N_F}$ is invariant under the change of variables $i\to L-i-1$ together with $k\to N_F-k+1$, which implies that the distribution for $\hat q_k$ can be obtained from $\hat q_{N_F-k + 1}$ via a spatial reflection around the center of the chain, as shown in Figs.~\ref{fig:fig1}(c). Moreover, unlike Eq.~\eqref{eq:Haaravg}, this distribution depends explicitly on $L$; however, for a fixed finite $k$ it still approaches a well defined finite distribution in the limit $L\to\infty$. For $k \propto L$, it once again has a width $\sim \sqrt{L}$, as shown in Figs.~\ref{fig:fig1}(c-d). Both the position of the peak and the width of the distribution are now functions of the filling fraction $\nu = N_F / L$. The position is $\bar{i} = k/\nu$, while the width goes to zero as $\nu\to 1$. In the thermodynamic limit, to leading order in $L$, one finds $p_{\nu L}(xL;\alpha L) \propto \exp\left[-L\left(h_2(\nu) - x h_2(\frac{\alpha}{x}) - (1-x) h_2(\frac{\nu-\alpha}{1-x})\right)\right]$, where the exponent is zero if $x = \alpha / \nu$ and negative otherwise. 

One can also calculate the variance, which has the same form as Eq.~\eqref{eq:HaarVar}, with $p_\text{Haar}$ replaced by $p_{N_F}$ and $3^L$ replaced by $\binom{L}{N_F}$, the dimension of the symmetry sector.

In principle, we could fix not only the particle number, but also the total magnetization $S^z_{\text{tot}}$. However, since the string operators $\hat{\mathcal{P}}_i^k$ do not depend on the local magnetization, the probability distribution $p_{N_F}(i;k)$ would remain the same for any $S^z_\text{tot}$. For the same reason, one would even have the same distribution for a random state within a sector with a fixed spin pattern.

A conceptual comparison between LIOMs and SLIOMs can be found in Table.~\ref{table:fig2}.
We emphasize that, although the two concepts play a similar role (providing labels for eigenstates and connected subspaces, respectively), there is also an important difference: LIOMs exist throughout the entire MBL phase and are only slightly modified by perturbations.
SLIOMs, on the other hand, are destroyed by generic perturbations (i.e., those not diagonal in the $S^z$ basis).

A similar comparison could be made between SLIOMs and conserved quantities of integrable models. 
We highlight that the two are rather different, SLIOMs can not be written as sums of local densities, unlike the conserved quantities in (Bethe ansatz) integrable models. Another difference is that SLIOMs can be used to block-diagonalize the Hamiltonian, while in interacting integrable systems, most conserved quantities have non-degenerate spectra, so diagonalizing them would be equivalent to fully diagonalizing the Hamiltonian~\cite{Pozsgay_2013,Ilievski_2016}.

A structure similar to the SLIOMs defined above arises in another strongly fragmented model, where the conserved quantities are harder to identify, as we shall see below in Sec.~\ref{sec:H3}.

\begin{table}
	\includegraphics[width = 1.\columnwidth]{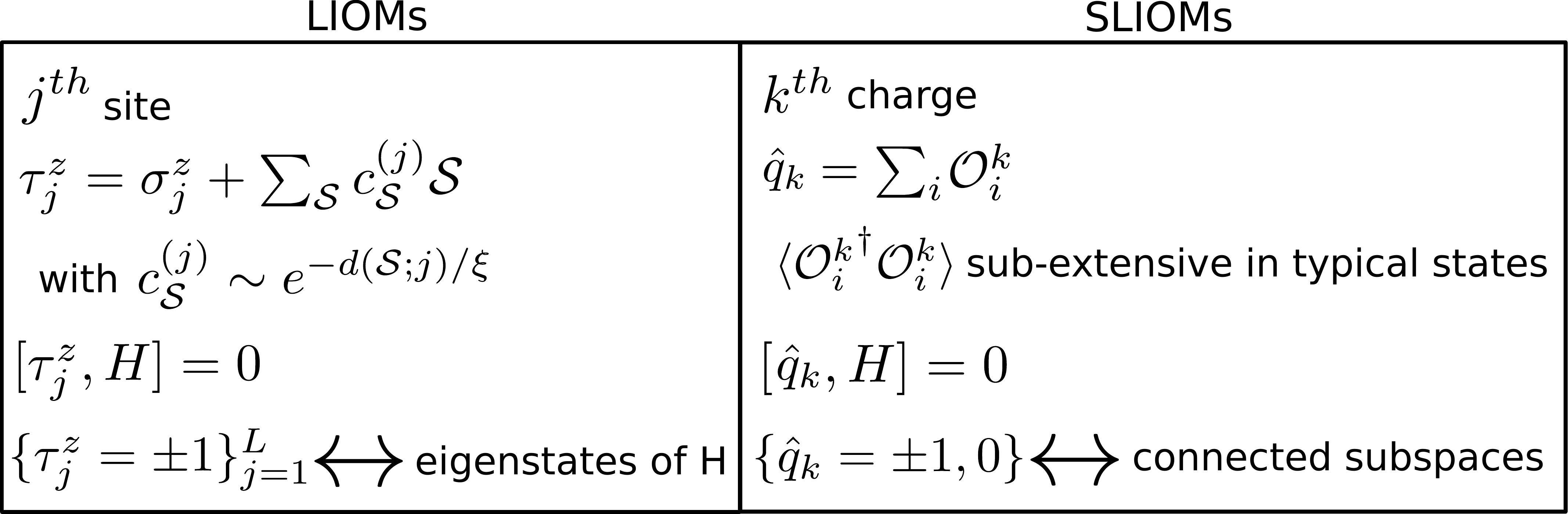}
	\caption{\textbf{Comparison between LIOMs and SLIOMs in the $t-J_z$ model.} While LIOMs label many-body eigenstates, SLIOMs label larger dimensional connected subspaces. LIOMs are localized around a given position in a state-independent way (operator strings $\mathcal{S}$ with support far from $i$ are exponentially suppressed). SLIOMs, on the other hand, are restricted to regions that depend on the state considered (e.g. its filling fraction for the $t-J_z$ model). Unlike LIOMs, which are always exponentially localized, the SLIOMs in the $t-J_z$ model are only partially localized with a width that is sub-extensive but infinite in the thermodynamic limit.}
	\label{table:fig2}
\end{table}
 
\subsection{Bulk vs boundary SLIOMs and their relationship to thermalization}\label{sec:tJz_thermal}

Having defined the conserved quantities that characterize the $t-J_z$ model and its fragmented Hilbert space, we now turn to the question of how these affect the dynamics, in particular whether they lead to a breakdown of thermalization.
As we shall see, the effect of SLIOMs is strongest near the boundary, where they lead to infinitely long coherence times, in complete analogy with the case of strong zero modes~\cite{Fendley_2012,Fendley_2016,Fendley16,Kemp17,Chetan17,Garrahan19}. 
In the bulk, we find that coherence times are finite in the thermodynamic limit, despite the presence of infinitely many conservation laws.
Nevertheless, even in the bulk, the SLIOMs lead to a weaker form of non-equilibration, wherein correlations remain trapped in a sub-extensive region, as well as to a violation of the eigenstate thermalization hypothesis within global symmetry sectors. 

\subsubsection{Bulk behavior}

\begin{figure}[t!]
	\includegraphics[width=1.\columnwidth]{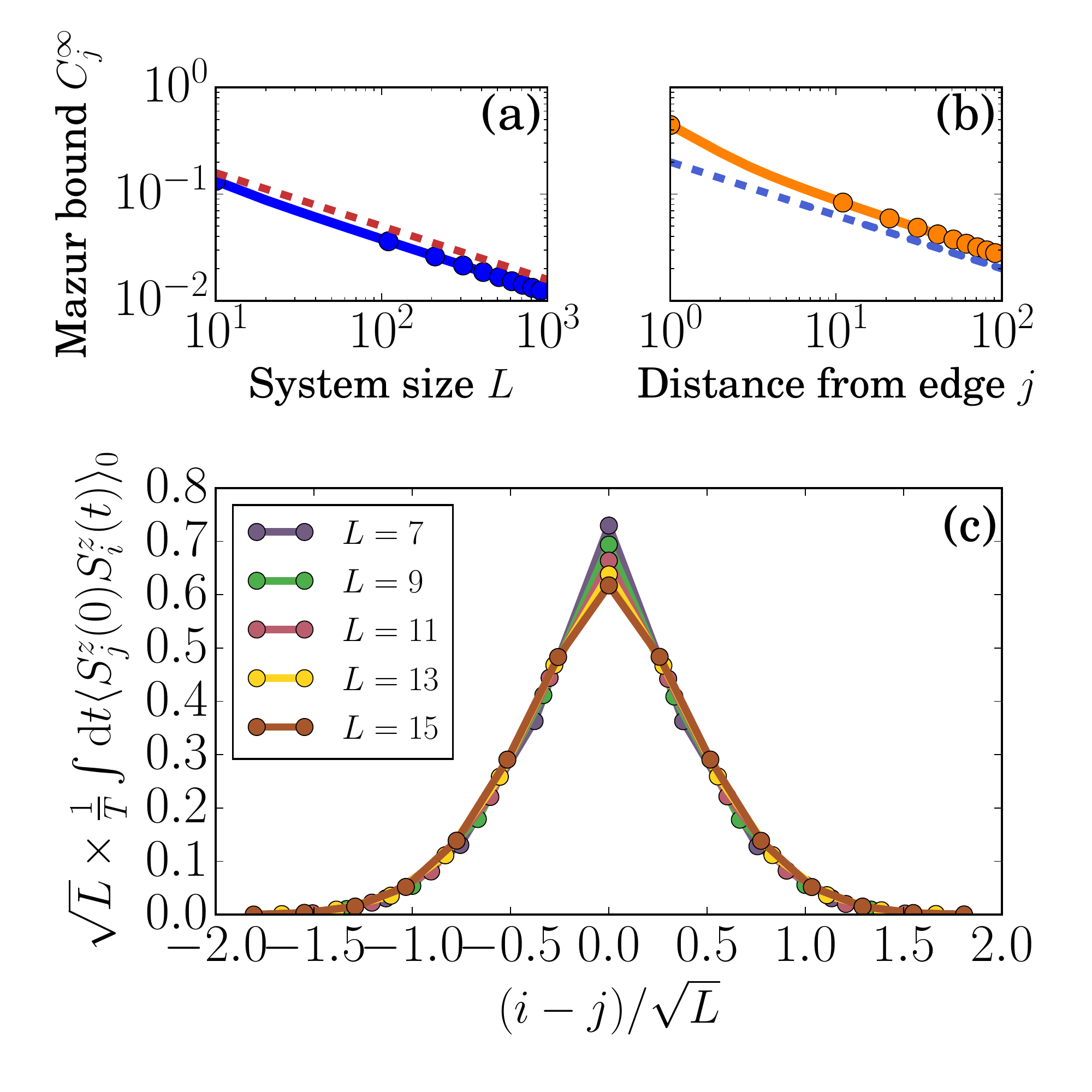}
	\caption{\textbf{Autocorrelations for the $t-J_z$ model in the bulk} (a) Mazur bound~\eqref{eq:mazur} on autocorrelations in the bulk, at $j=L/2$, decays as $\propto L^{-1/2}$  as a function of the system size $L$. (b) The same bound, shown for a fixed $L=600$, decays as $\propto j^{-1/2}$ as a function of the distance $j$ from the boundary. (c) The long-time average of spatially resolved correlations, computed numerically for small chains (and averaged between times $t=50$ and $100$), shows a persistent peak, instead of the complete spreading expected from thermalization.}
	\label{fig:tJz_autocorr}
\end{figure}

A natural question to ask regarding thermalization is whether the presence of an extensive number of SLIOMs manifests itself in infinite autocorrelation times, as is the case in MBL. A way to gain insight into this question is by considering Mazur's inequality~\cite{Mazur69,Suzuki71,Caux10}, which provides a lower bound on the time-averaged autocorrelation of an observable based on its overlap with the conserved quantities. Focusing on a single-site $S_j^z$ operator, and considering only the SLIOMs $\hat q_k$, the inequality in our case reads 
\begin{multline}\label{eq:mazur}
\lim_{T\to\infty} \frac{1}{T} \int\text{d}t \, \braket{S_j^z(t)S_j^z}_{\beta=0} \geq \sum_k \frac{|\braket{S_j^z\hat{q}_k}_{\beta=0}|^2}{\braket{\hat{q}_k^2}_{\beta=0}} = \\
= \sum_k \frac{\left[ 3^{-j} 2^{k} \binom{j-1}{k-1} \right]^2}{1 - 3^{-L} \sum_{N_F=0}^{k-1}2^{N_F} \binom{L}{N_F}} \equiv C_j^z(\infty),
\end{multline}
where $\braket{A}_{\beta=0} \equiv \text{tr}(A) / 3^{L}$ is the infinite temperature average, and the denominator in the last expression is the probability of having at least $k$ particles in the system. If the expression on the right hand side of this inequality was finite in the limit $L\to\infty$, it would imply infinitely long coherence times. Instead, evaluating it for a bulk observable, $j \propto L$, one finds that it decays with system size as $L^{-1/2}$, as shown by Fig.~\ref{fig:tJz_autocorr}(a). This implies that the conservation laws $\{ \hat q_k\}$ are not sufficient to prevent the autocorrelation from decaying to zero at long times.

Even though the bound vanishes in the thermodynamic limit, it nevertheless implies anomalous dynamics. 
For a conserved density like $S_j^z$, one expects the spatially resolved autocorrelation $\braket{S_j^z(t)S_i^z}$ to eventually spread out over the whole system and thus become $O(1/L)$ for all $i$. 
However, in our case the lower bound $\sim L^{-1/2} \gg L^{-1}$ implies that this cannot be the case, and instead suggests that the charge remains trapped within a much smaller region of size $O(L^{1/2})$.  
This can be understood from the distribution of the conserved quantities in Fig.~\ref{fig:fig1}, which we discussed in the previous section. 
In particular, note that the infinite temperature overlap $\avg{S_j^z\hat{q}_k}_{\beta=0} $ is proportional to the value of the probability distribution $p_\text{Haar}(j;k)$ in Eq.~\eqref{eq:Haaravg}, since $\text{tr}(S_j^z\hat{q}_k)=\text{tr}({\mathcal{O}^k_j}^\dagger \mathcal{O}^k_j)$.
As we saw above, SLIOMs in the bulk have a width $\propto L^{1/2}$. 
Therefore, a given $S_j^z$ overlaps significantly with only $O(L^{1/2})$ different conserved quantities $\hat q_k$, and these define the region in which the charge can spread out. 
This conclusion is supported by numerical results on the spatially resolved correlator $\braket{S_j^z(t)S_i^z}_{\beta=0}$ at long times for small chains, as shown by Fig.~\ref{fig:tJz_autocorr}(c).
These results suggest a scaling $\braket{S_j^z(t) S_i^z}_{\beta=0} \approx \frac{1}{\sqrt{L}} f(\frac{i-j}{\sqrt{L}})$ in the limit of large $L$.

While autocorrelations in the bulk thus decay to zero at long times in the thermodynamic limit (albeit in an anomalous manner), this does not imply that the system thermalizes. 
Indeed, an initial product state in the fermion occupation basis would clearly not relax to a thermal state solely specified by the global conserved quantities $H_{t-J_z}$, $N_F$ and $S^z_\text{tot}$.
In particular, since each sector with a fixed pattern of spins is effectively a chain of spinless fermions with 2 possible states per site, time evolving from such an initial state will result in half-chain entanglement entropies at most $\frac{L}{2}\ln{2}$, much smaller than the entropy of a chain with 3-dimensional local Hilbert space at (or close to) infinite temperature ($\frac{L}{2}\ln{3}$). 
One could say that each of these initial states thermalizes with respect to the associated effective spinless fermion Hamiltonian, i.e. the $t-J_z$ Hamiltonian projected to a given connected sector with a fixed value of the SLIOMs.
Note, however, that this effective Hamiltonian is non-local: to know the sign of the interaction between a given pair of (spinless) fermions, one in principle has to know the entire spin pattern in the original variables.

\begin{figure}
	\includegraphics[width=0.75\columnwidth]{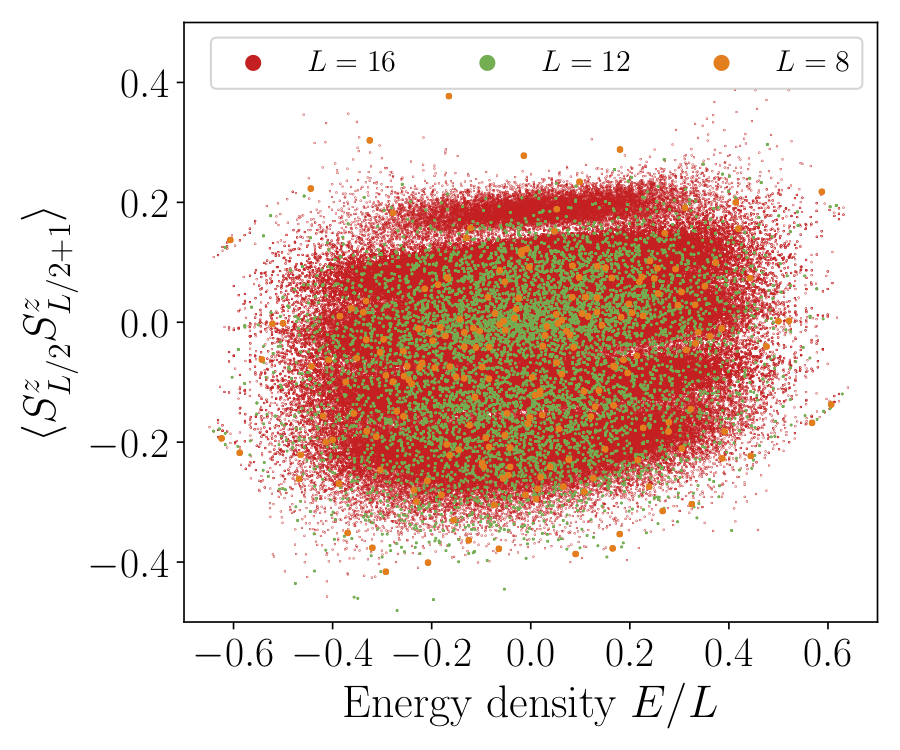}
	\caption{\textbf{Diagonal matrix elements in the $t-J_z$ model.}  Expectation value of the average nearest neighbor antiferromagnetic correlations in eigenstates of $H_{t-J_z}$ with global quantum numbers $N_F = L/2$ and $\sum_j S_j^z = 0$, and open boundary conditions. For the system sizes shown ($L=8,12,16$), the distribution becomes \emph{wider} with increasing system size, while asymptotically it is expected to narrow as $\sim L^{-1/4}$. This is a consequence of the strong fragmentation labeled by the SLIOMs, and is in contrast with ETH, which predicts an exponentially narrow distribution.
	}
	\label{fig:eth_open}
\end{figure}

This sensitivity to initial conditions, due to the presence of bulk SLIOMs, is also reflected in the properties of the eigenstates of $H_{t-J_z}$.
As the above argument shows, they have at most $\frac{L}{2}\ln{2}$ entanglement (for a half chain), much smaller than a generic Hamiltonian with 3 states per site would have in the middle of the spectrum.
Moreover, due to the strong fragmentation of the Hilbert space, different eigenstates at the same energy density, and with the same global quantum numbers $N_F$ and $S^z_\text{tot}$, can have very different expectation values for simple local observables. 
This is trivially true for the symmetry sectors with $N_F = L$, where all states are completely frozen, but it in fact holds more generally.
To confirm this, we consider the global symmetry sector with $N_F = L/2$ and $S_{\text{tot}}^z = 0$, and numerically evaluate the eigenstate expectation values of the observable $S_{L/2}^z S_{L/2+1}^z$. 
We find (see Fig.~\ref{fig:eth_open}) that the expectation values of this operator have a wide distribution over different eigenstates.
Approximating the eigenstates by an equal weight superposition of all possible hole positions with a given spin pattern, on the other hand, suggests that in fact there is a very slow narrowing of this distribution, with the width scaling as $L^{-1/4}$ in the thermodynamic limit as obtained from Monte Carlo simulation~\cite{JF}.
This slow algebraic narrowing should be contrasted with the ETH ansatz, which predicts an exponentially narrow distribution.
In fact, the $L^{-1/4}$ scaling is even slower than the case of integrable systems, which typically have a width $\sim L^{-1/2}$~\cite{Vidmar_2016,Vidmar_2019_1,Vidmar_2019_2}\footnote{In general, the eigenstate-to-eigenstate fluctuations of a local observable in any generic translation invariant system should decay at least as fast as $\sim L^{-1/2}$~\cite{Biroli2010,MoriWeakETH}.}; this difference is consistent with our picture of SLIOMs wherein the local observable only `sees' an $O(\sqrt{L})$ part of the system.

From these results, we conclude that if one considers only the global $(N_F,S^z_\text{tot})$ symmetry sector, without resolving the additional non-local symmetries, then the diagonal matrix elements of local observables violate ETH. 
This can be understood as follows: each connected sector has a different `embedded' Hamiltonian, depending on the spin pattern, and the properties of the associated eigenstates can therefore differ from sector to sector.
Note that this situation is different from the case of more commonly occurring non-local symmetries, such as spin-flips or lattice translations, which \emph{do not} lead to distinct distributions of diagonal matrix elements ~\cite{Sorg2014,Rigol2DIsing,Shiraishi17Comment,Shiraishi17Reply}\footnote{If this was not the case, systems with a discrete symmetry would not thermalize, since typical initial states do not have a sharply defined value of these conserved quantities.}. 
Of course one can instead consider only eigenstates within a given sector, in which case ETH is fulfilled for typical spin patterns (with the exception of a few integrable sectors, which we discuss below). 
Note, however, that this requires fixing an extensively large number of non-local symmetries (the SLIOMs)\footnote{We note here that not all different spin patterns give rise to distinct distributions of diagonal matrix elements. We leave it as an open question to identify exactly which combinations of the SLIOMs would need to be fixed to obtain a set of eigenstates that obey ETH.}, making difficult to meaningfully compare different system sizes.
In this sense, our case is similar to that of integrable models, where one usually considers matrix elements without resolving all the extensively many conserved quantities, and finds a similarly slow, algebraic decay of their fluctuations with system size~\cite{Vidmar_2016,Vidmar_2019_1,Vidmar_2019_2}.

So far we discussed the non-ergodicity originating from the fragmented Hilbert space, whose components are labelled by the SLIOMs. 
Our conclusions about the lack of thermalization therefore apply independently of the structure of the Hamiltonian \emph{inside} the connected blocks. 
For the $t-J_z$ Hamiltonian~\eqref{eq:tjz} it turns out that there is some additional structure for sectors with a completely ferromagnetic or completely antiferromagnetic spin pattern.
These can be mapped~\cite{Zhang97} onto a spin-1/2 XXZ Heisenberg chain (with anisotropy $\Delta > 0$ and $\Delta < 0$, respectively), which is quantum integrable. 
Most of the other sectors, on the other hand, show random matrix level statistics, signalling quantum chaotic behavior.
The integrability of the FM and AFM sectors could also be broken by additional perturbations that are diagonal in the $S^z$ basis (e.g. a staggered field). 
These commute with all the SLIOMs, and therefore do not change our conclusions about the overall non-ergodicity of the model. 

\subsubsection{Statistically localized strong zero modes}\label{sec:zero_modes}

It is worthwhile to consider separately those constants of motion $\hat{q}_k$ that are localized at the boundary of an open chain. 
In this case $k$ does not scale with the system size and therefore its distribution $p_\text{Haar}(i;k)$ remains finite in the thermodynamic limit. 
Consequently, one expects that an observable near the boundary has finite overlap with these SLIOMs and, under time evolution, a non-vanishing fraction of it would remain localized in a finite region near the boundary.
Indeed, computing the lower bound from Eq.~\eqref{eq:mazur} for a position $j$ that does \emph{not} scale with $L$, one finds that it remains finite in the limit $L \to \infty$. 
The bound is largest at the boundary, $j=1$, where it takes the value $4/9$, and decays away from the boundary as $j^{-1/2}$.
This is shown in Fig.~\ref{fig:tJz_autocorr}(b).
Obviously, the same holds near the right edge, when $j$ is replaced by $L+1-j$.
Therefore, at the boundaries the SLIOMs imply a much stronger breaking of thermalization, resulting in infinite coherence times.

In fact, in order to derive infinite coherence times at the edge, one does not need infinitely many SLIOMs, it is sufficient to consider just \emph{one}. In particular let us take the spin of the leftmost fermion,
\begin{equation}\label{eq:ql_def}
\hat{q}_\ell \equiv \sum_i {\Big (} \prod_{j < i}(1 - \tilde{n}_j) {\Big )} S_i^z,
\end{equation}
which is equivalent to $\hat{q}_{k=1}$ in the above definition, with the projection taking a particularly simple form $\hat{\mathcal{P}}^1_i = \prod_{j < i}(1 - \tilde{n}_j) \tilde{n}_i$, using the local constrained fermion density $\tilde{n}_j=\tilde{n}_{j,\uparrow}+\tilde{n}_{j,\downarrow}$. 
There is another similar operator localized near the right edge
\begin{equation}\label{eq:qr_def}
\hat{q}_r \equiv \sum_i S_i^z {\Big (} \prod_{j > i}(1 - \tilde{n}_j) {\Big )}.
\end{equation}
A reason to highlight these \emph{boundary SLIOMs} is that they already lead to infinite coherence times at the two edges, without having to consider the other conserved quantities. 

\begin{figure}
	\includegraphics[width =1.\columnwidth]{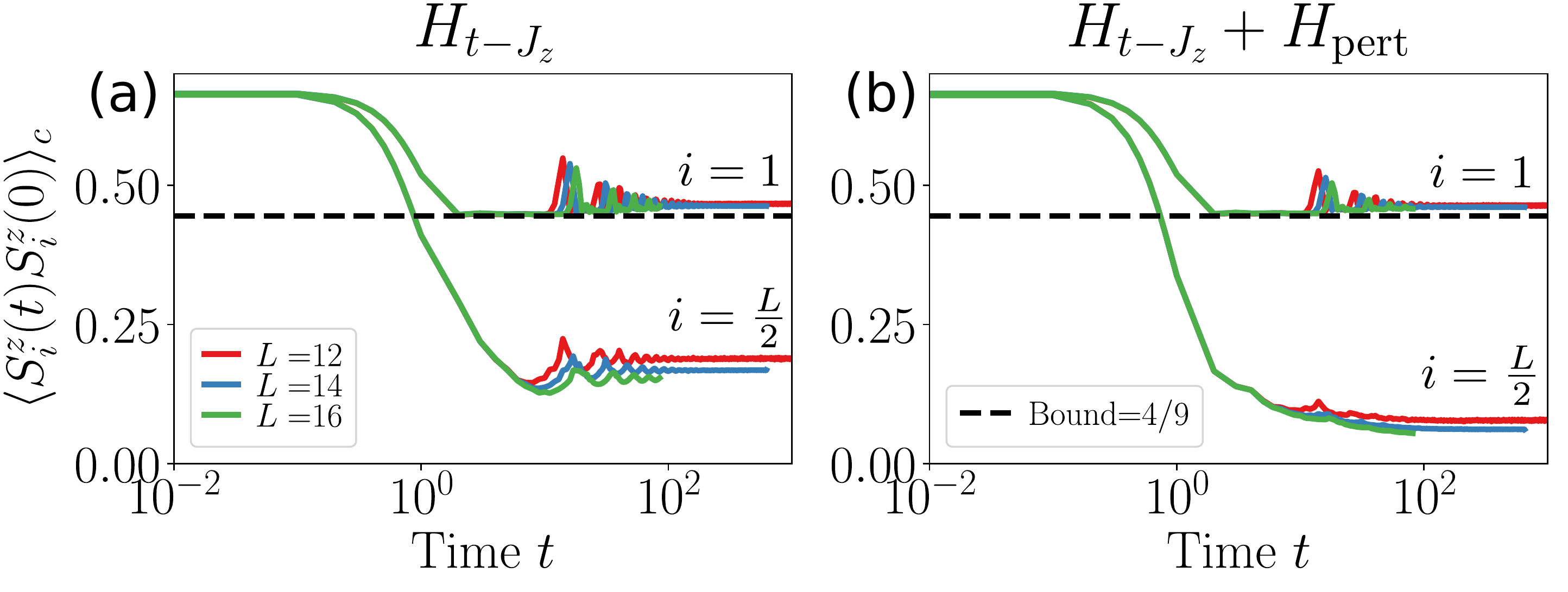}
	\caption{\textbf{Bulk vs edge autocorrelations.} Connected infinite temperature autocorrelation function for the center site $i=L/2$ and at the left boundary $i=1$ for system sizes $L=11,13,15$. (a) In the $t-J_z$ model (Eq.~\eqref{eq:tjz}), which conserves both bulk and boundary SLIOMs $\hat q_k$, the edge autocorrelator shows infinite coherence times while in the bulk it decays to a value $\propto L^{-1/2}$, which is anomalously large but vanishing in the thermodynamic limit. (b) Once the perturbation \eqref{eq:Hpert} is added, SLIOMs in the bulk are broken and the bulk autocorrelations decay to the value $\propto 1/L$ expected for thermalizing systems. The boundary SLIOMs $\hat q_\ell, \hat q_r$, on the other hand, are still conserved, leading to a finite long-time value for autocorrelations at the edge, well approximated by the analytical lower bound (dashed horizontal line).}
	\label{fig:fig2}
\end{figure}

Once more, we make use of Mazur's inequality. The conservation law $[\hat{q}_\ell,H] = 0$ implies that
\begin{equation}\label{eq:mazur_boundary}
\lim_{T\to\infty} \frac{1}{T} \int\text{d}t \,\braket{S_j^z(t)S_j^z}_{\beta=0} \geq \frac{|\braket{S_j^z\hat{q}_\ell}_{\beta=0}|^2}{\braket{\hat{q}_\ell^2}_{\beta=0}} = \frac{4}{9^j}.
\end{equation}
In evaluating the right hand side we used the fact that $3^{-L}\text{tr}(S_j^z\hat{q}_\ell) =2/3^j$ as given by Eq.~\eqref{eq:Haaravg}, and $\hat{q}_\ell^2 = 1 - \hat{P}_\text{empty}$ where $\hat{P}_\text{empty}$ is a rank 1 projector onto the completely empty state. 
One can do the same calculation near the right boundary, for $S_{L+1-j}^z$, using the conservation of $\hat{q}_r$, which leads to the lower bound $4/9^{L+1-j}$. 

While this result is weaker than the one taking all the $\hat{q}_k$ into account (it decays exponentially, rather than algebraically, towards the bulk), it follows from much weaker conditions.
This implies that it is possible to add perturbations to the Hamiltonian that destroy the strong fragmentation in the bulk, but nevertheless lead to non-thermalizing dynamics at the edge. 
A simple example of such a perturbation is
\begin{equation} \label{eq:Hpert}
H_\text{pert} = \sum_{i=2}^{L-2} \tilde{n}_{i-1} (S_i^x S_{i+1}^x + S_i^y S_{i+1}^y) \tilde{n}_{i+2},
\end{equation}
which allows spins to flip-flop, but only if both neighboring sites are occupied by a fermion. 
Therefore, this perturbation no longer conserves the spin pattern, but it still commutes with the two boundary SLIOMS, $\hat q_{\ell,r}$.

As a consequence, the bound~\eqref{eq:mazur_boundary}, evaluated at the boundaries, applies to the perturbed Hamiltonian $H_{t-J_z} + \lambda H_\text{pert}$, despite that it is now completely thermalizing in the bulk.
As shown in Fig.~\ref{fig:fig2}, the lower bound derived from Mazur's inequality appears to be tight for the boundary autocorrelation, while the bulk autocorrelation in the perturbed system now decays to an $O(1/L)$ value, as expected for a thermalizing system.

The appearance of infinitely long coherence times at the boundaries is strongly reminiscent to the case of strong edge modes previously discussed in the literature~\cite{Fendley_2012,Fendley_2016,Fendley16,Kemp17,Chetan17,Garrahan19}. The operators $\hat{q}_{\ell,r}$ play the same role as the strong zero modes (SZM), whose presence prevents boundary operators from thermalizing. The differences are twofold: i) Our boundary modes are only statistically localized, in the sense defined above, unlike the usual SZM which are localized in an operator sense. ii) On the other hand, in our case $\hat{q}_{\ell,r}$ commute \emph{exactly} with the Hamiltonian for arbitrary system sizes, unlike the strong zero modes which only commute up to $O(e^{-L})$ corrections. One can find a comparison between SZMs and boundary SLIOMs in Table.~\ref{table:fig3}.

\begin{table}
	\includegraphics[width = 1.\columnwidth]{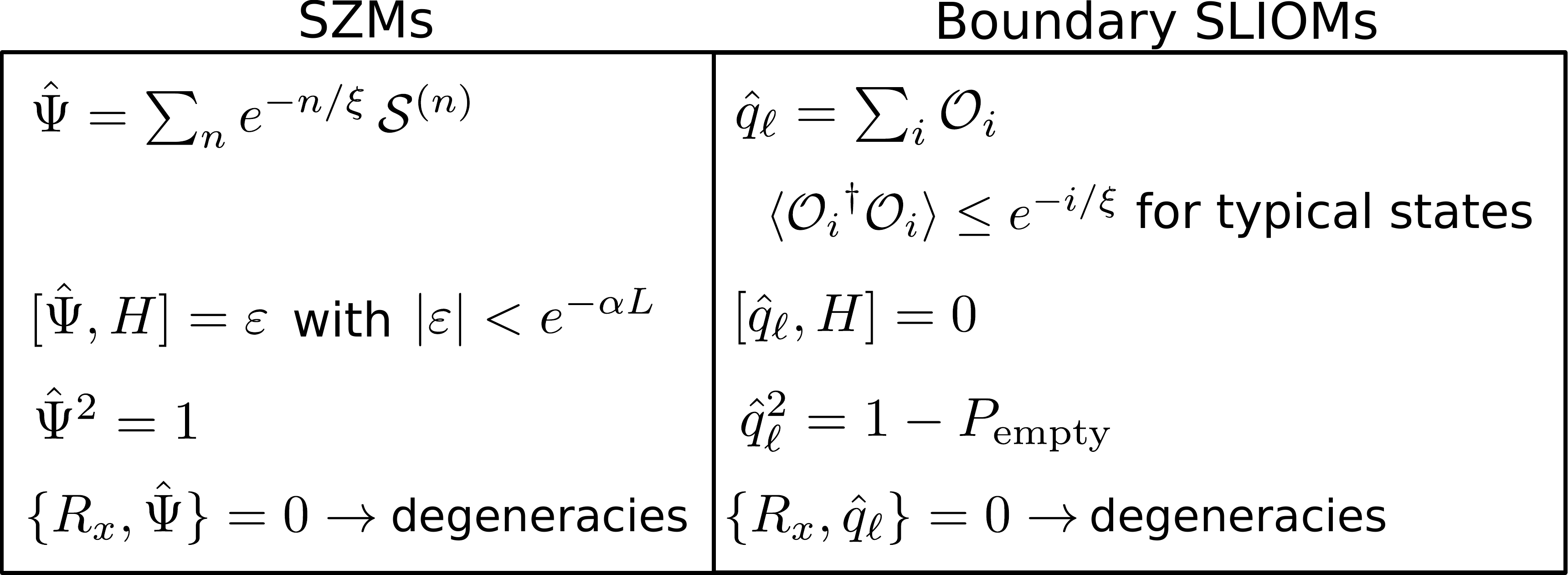}
	\caption{\textbf{Comparison between boundary SLIOMs and strong zero modes (SZM).} The SZM is a sum of string operators $\mathcal{S}$ (e.g. Jordan-Wigner strings), ending at distance $n$ from the edge, with coefficient decaying exponentially with $n$. For the boundary SLIOMs, on the other hand, localization appears upon taking the expectation value in typical states with finite particles density. While SZM are usually only conserved in the thermodynamic limit, the SLIOMs are exact integrals of motion at any finite size $L$. The existence of an additional symmetry, in this case
	$R_x=\prod_j \exp\left(i\pi \frac{S_j^x}{2}\right)$,
	anti-commuting with the SLIOMs or edge modes, implies degeneracies throughout the many-body spectrum. Majorana SZM square to $1$, while in our case $\hat{q}_{\ell,r}^2$ is $1$ everywhere except in a particular one-dimensional subspace (i.e., a state with no particles).}
	\label{table:fig3}
\end{table}

The fact that $H = H_{t-J_z} + \lambda H_\text{pert}$ commutes with the two edge mode operators means that it can be decomposed into four blocks, according to the spin of the left- and rightmost fermions, written formally as $H = H_{\uparrow\uparrow} \oplus H_{\downarrow\downarrow} \oplus H_{\uparrow\downarrow} \oplus H_{\downarrow\uparrow}$ (excluding the empty state).  
Eigenstates can therefore be labeled by the left- and rightmost spins.
In the presence of additional symmetries, not commuting with $\hat q_\ell$ and $\hat q_r$, this implies degeneracies in the energy spectrum at all energies, just as in the case of usual strong edge modes. 
In particular, $H_{t-J_z}$ and $H_\text{pert}$ are both invariant under flipping all spins simultaneously i.e.,
$R_x=\prod_j \exp\left(i\pi \frac{S_j^x}{2}\right)$.
%
This operator flips the eigenvalues of both $\hat{q}_\ell$ and $\hat{q}_r$, and therefore interchanges the blocks $H_{\uparrow\uparrow} \leftrightarrow H_{\downarrow\downarrow}$ and $H_{\uparrow\downarrow} \leftrightarrow H_{\downarrow\uparrow}$.
This implies that the spectrum is at least 2-fold degenerate everywhere; since the Hamiltonian commutes with $\hat{q}_\ell, \hat{q}_r, R_x$ at any finite size, this degeneracy is \emph{exact}.

Given the presence of such edge modes throughout the entire spectrum, it is natural to ask whether the ground state of $H_{t-J_z}$ is in a topological phase. This is in fact not as obvious as it might seem, for two reasons: firstly, the type of edge mode operators we have discussed are known to also emerge in symmetry-breaking phases\footnote{One can think of the edge mode as measuring a spontaneous boundary magnetization. In the absence of a bulk magnetization, this implies symmetry protected topological phases. However, if the bulk is magnetized, the edge magnetization is simply picking this up.}---indeed this happens in the large $J_z$ limit---and secondly, we have already noted that we can essentially trivialize the bulk whilst preserving the edge mode (with perturbations of the type in Eq.~\eqref{eq:Hpert}), in which case the ground state can be trivial in the bulk\footnote{This would mean that the edge mode is not stabilized by symmetry alone but requires the boundary SLIOM.}.

Nevertheless, it turns out that the ground state is in a topologically non-trivial phase. This is all the more intriguing when one observes that the $t-J_z$ model, as defined in Eq.~\eqref{eq:tjz}, is \emph{gapless} for $0 < J_z < t$ (to wit, we consider $J_z = t/4$), whereas (symmetry-protected) topological phases are usually gapped. Recently, frameworks for gapless topological phases have been introduced ~\cite{Vasseur17,Verresen19}. In fact, the ground state of the $t-J_z$ model appeared as a particular example of a (topologically non-trivial) symmetry-enriched critical point in Sec.~VII.A of Ref.~\onlinecite{Verresen19}; there it was discussed in the formulation as a spin-$1$ chain, with the Hamiltonian arising as the simplified version of the gapless Haldane phase first introduced in Ref.~\onlinecite{Kestner11} protected by $\mathbb Z_2 \times \mathbb Z_2$. Interestingly, the topologically non-trivial nature of the gapless $t-J_z$ model was noted over two decades ago in Ref.~\onlinecite{Zhang97} in terms of a hidden antiferromagnetic order, although the twofold ground state degeneracy was not observed. As we have noted above, this twofold degeneracy is \emph{exact} in this case.
The $\mathbb Z_2 \times \mathbb Z_2$ symmetry group of the spin-$1$ chain studied in Ref.~\onlinecite{Verresen19}, maps to the fermionic parity and $U=\prod_i U_i$ with $U_i\equiv \ket{0}\bra{0} -\ket{\uparrow}\bra{\downarrow} - \ket{\downarrow}\bra{\uparrow}$ in the fermionic formulation~\cite{Kestner11}.
Our above definition of $R_x$ replaces this second $\mathbb Z_2$ by a $\mathbb Z_4$ symmetry group.

If we add an \emph{arbitrary}\footnote{We note that the edge mode is stable against opening up a bulk gap, as discussed in Ref.~\onlinecite{Verresen19}} perturbation (breaking the bulk \emph{and} edge SLIOMs) that preserves either of the above symmetry groups, then this twofold degeneracy\footnote{If the perturbation drives us into a gapped symmetry-breaking phase, the total degeneracy is twofold; if we are driven to a gapped symmetry-protected topological phase, the degeneracy becomes fourfold due to the finite correlation length decoupling the two edges.} would only persist at low energies and would acquire an exponentially small finite-size splitting, per the arguments in Refs.~\onlinecite{Vasseur17,Verresen19}.

\subsection{Experimental realization}
Ultracold atoms in a shallow optical lattice that are optically dressed with a Rydberg state, realize a variant of the $t-J_z$ model of Eq.~\eqref{eq:tjz}~\cite{ZeiherNatPhys2016,ZeiherPRX2017}. The Hamiltonian of the Rydberg system is given by
\begin{align} \label{eq:Rydberg} \nonumber
H_\text{Rydberg} &= -t \sum_{\substack{i, \sigma}} (\tilde{c}_{i,\sigma} \tilde{c}_{i+1,\sigma}^\dagger + \text{H.c.} ) \\ &+ \sum_{i\neq j} \frac{U_0/8}{1+(r_{ij}/R_c)^6} \ket{\uparrow_i\uparrow_j} \bra{\uparrow_i\uparrow_j}.
\end{align}
Here, the first term describes the hopping of the atoms, which possess two internal states, $\ket{\downarrow}$ and $\ket{\uparrow}$, in a one-dimensional optical lattice. The atoms can have either fermionic or bosonic statistics, as for the latter a hard-core constraint is typically enforced due to the strong Rydberg interactions. The interaction potential is of strength $U_0=\Omega^4/8|\Delta|^3$ and has a cutoff at $R_c=2\Delta$, where $\Omega$ is the Rabi frequency and $\Delta$ the detuning from the Rydberg sate~\cite{HenkelPRL2010}. This potential can be adjusted such that it effectively acts only on nearest-neighbor sites with some strength $J_z$~\cite{ZeiherPRX2017}. 
Since the two Hamiltonians only differ by diagonal terms, our results for SLIOMs in the $t-J_z$ model \eqref{eq:tjz} carry directly over to the Rydberg system. 

Moreover, we can partially break the structure of the SLIOMs in the bulk by engineering for the Rydberg system a perturbation in the spirit of the one in \eqref{eq:Hpert}. In particular, when coupling the two internal states, $\ket{\downarrow}$ and $\ket{\uparrow}$, with a global microwave of strength $\Omega_\text{mw} \ll J_z$ that is blue detuned by $2J_z$ from the atomic transition, an effective coupling of the form $\sum_i (\ket{\uparrow}\bra{\uparrow})_{i-1} S^x_i (\ket{\uparrow}\bra{\uparrow})_{i+1}$ is generated in the rotating frame of the Rydberg interaction~\cite{LesanovskiPRL2011,WintermantelArxiv2019}. One can realize this perturbation in addition to the Rydberg interaction, for example by pulsing the microwave drive. This perturbation does not preserve the total charge but nevertheless has an effect similar to \eqref{eq:Hpert}, destroying the SLIOMs in the bulk while maintaining them at the boundary. 

Note that the systems considered in this section are different from those in Eqs.~\eqref{eq:tjz} and~\eqref{eq:Hpert}, in that they are not invariant under the symmetry transformation
$R_x=\prod_j \exp\left(i\pi \frac{S_j^x}{2}\right)$.
Therefore, these models do not show the exact twofold degeneracy of the spectrum previously discussed. Nevertheless, they exhibit the same physical phenomena with respect to thermalization as the ones discussed above.

\section{Dipole-conserving Hamiltonian $H_3$}\label{sec:H3}

The example of the $t-J_z$ model may seem somewhat trivial, since the connected components of the Hilbert space can be easily read off from the Hamiltonian. Here we show that the same general concept of statistically localized integrals of motion applies to a more complicated Hamiltonian~\cite{Sala19}. However, we will also highlight some differences between the two cases. In particular, while in the $t-J_z$ model the starting point of the identification of sectors was related to the number of fermions, a usual U(1) symmetry, in the case discussed below the analogous quantity (the number of objects whose pattern is conserved) is already non-local in terms of the physical degrees of freedom. Moreover, while $H_{t-J_z}$ only had partially localized conserved quantities, the model we consider in the following also exhibits SLIOMs that are statistically localized to finite regions, leading to infinite coherence times even in the bulk.

The system we consider is a spin-1 chain, with a 3-site Hamiltonian that, apart from the total $S^z$ component $Q=\sum_jS_j^z$ (`charge'), also conserves its associated dipole moment, $P \equiv \sum_j j S_j^z$. It reads
\begin{equation}\label{eq:H3}
H_3 = -\sum_j S^+_{j-1} (S^-_j)^2 S^+_{j+1} + \text{H.c.}
\end{equation}
In the following we will denote the three on-site eigenstates of $S_j^z$ by $\ket{+},\ket{-},\ket{0}$ (corresponding to eigenvalues $+1,-1,0$), and refer to them, respectively, as a positive charge, a negative charge, and an empty site. 
In the following, we take open boundary conditions.
Such dipole-conserving Hamiltonians appear as effective descriptions in a variety of settings, such as fracton systems~\cite{PretkoSub,Pai18}, the quantum Hall effect~\cite{RezayiHaldane,Bergholtz08,BERGHOLTZ2011,Nakamura2012, Sanjay19}, and for charged particles in a strong electric field~\cite{Refael18,Stark18}.

The Hamiltonian~\eqref{eq:H3} was shown to be non-ergodic~\cite{Sala19}, due to the strong fragmentation of the Hilbert space in the local $S^z$ basis into exponentially many invariant subspaces of many different sizes.
However, finding a set of labels that characterize these sectors was left open. 
Here we remedy this, constructing a full set of conserved quantities which completely characterize the block structure of $H_3$ in the local $S^z$-basis. Moreover, we show that they follow the recipe of statistically localized operators outlined above, but have a much richer structure than the $t-J_z$ model described in the previous section.
This additional structure accounts for the fact that $H_3$ has a much broader distribution of the sizes of connected sectors and a localized behavior \emph{in the bulk} in the form of infinite autocorrelation times, a feature not present in $H_{t-J_z}$.

\subsection{Mapping to bond spins and defects}

In order to identify the structure of connected sectors, it is useful to rewrite the dynamics in terms of a new set of variables. 
These new variables consist of two different types of degrees of freedom: spin-1/2 variables associated to the \emph{bonds} of the original chain---with corresponding Pauli operators denoted by $\sigma_{j,j+1}^{x,y,z}$ on the bond $(j,j+1)$---and hard-core particles living on the sites, which we will refer to as \emph{defects}.
To get a one-to-one mapping between basis states in the original $S_j^z$ basis and the new variables, we require the spins on the two bonds surrounding a defect to be aligned. 
Introducing the defect occupation number operator $n^d_j$ on site $j$, we can write this requirement formally as $\sigma_{j-1,j}^z n_j^d\ket{\psi} = \sigma_{j,j+1}^zn_j^d\ket{\psi}$ for any physical state $\ket{\psi}$.
With this constraint, the two Hilbert spaces match up and we get a mapping between basis states in the original $S_j^z$ basis and the new variables, as we now explain.

\begin{figure}[t!]
	\centering
	\includegraphics[width = 1.\columnwidth]{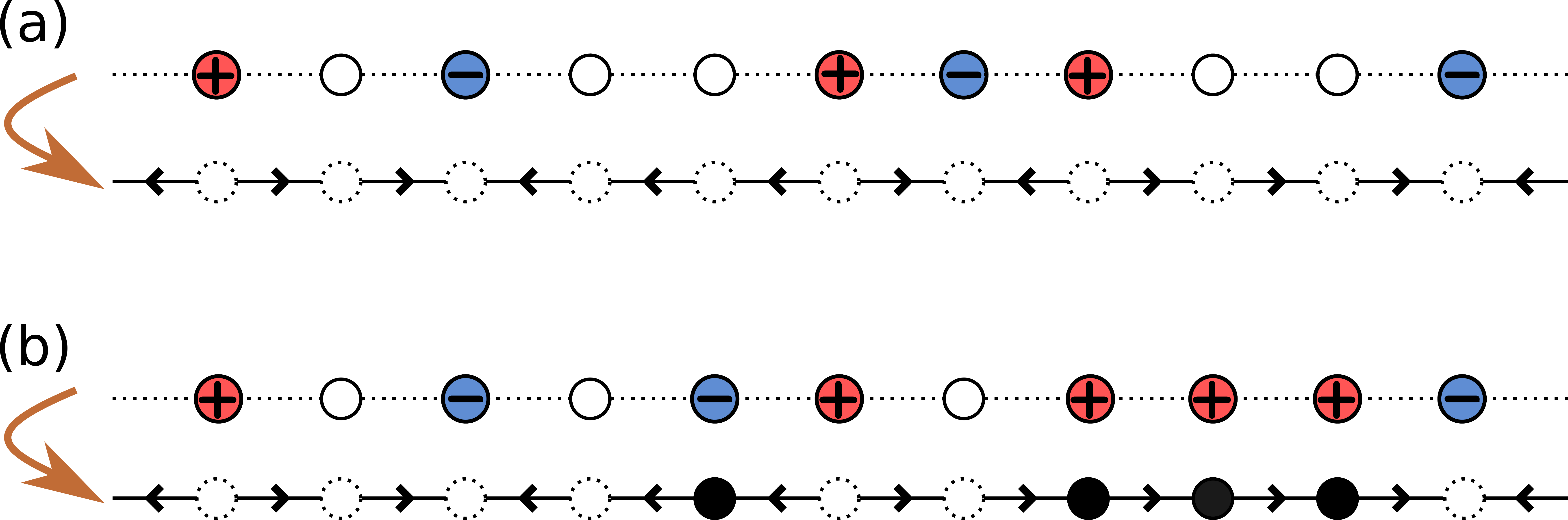}
	\caption{\textbf{Mapping from spin-1 chain to bond spins and defects.} (a) A charge configuration with alternating signs can be mapped to spin-1/2 variables on the bonds. (b) For a generic configuration, one also has to introduce defects, living on sites, whenever a charge would violate the rule of alternating signs. Note that defects with neighboring bond spins pointing to the right (left) correspond to positive (negative) charges in the original.}
	\label{fig:mapping_sketch}
\end{figure}

In order to understand how the mapping works, let us start considering those configurations of the original variables, which obey the following rule: \emph{subsequent charges---ignoring empty sites in-between---have alternating signs}\footnote{In other words, these are the set of states that have perfect antiferromagnetic ordering after eliminating the intermediate empty sites.}.
We can map a configuration of charges satisfying this rule to a configuration of bond spins with the following convention: we represent spins as pointing left ($\leftarrow$) or right ($\rightarrow$) and map each $(+)$-charge to a domain wall of type $\leftarrow \rightarrow$, and each $(-)$-charge to a domain wall of type $\rightarrow\leftarrow$, as shown in the example of Fig.~\ref{fig:mapping_sketch}(a).
To account for all configurations, we need to include two additional auxiliary bonds ($L+1$ bonds in total), at the left and right ends of the chain, whose spin configuration is fixed by the sign of the left- and rightmost charges respectively.
A way of visualizing the mapping is to think of the bond spins as an electric field, emanating from positive charges and ending at negative charges, satisfying Gauss's law, $\sigma_{j,j+1}^z - \sigma_{j-1,j}^z = 2 S_j^z$, where the operator $S_j^z$ measures the on-site charge in the original (spin-1) variables. 
The rule of alternating signs ensures that this prescription is consistent within the spin-1/2 representation on the bonds. 

The mapping to bond spins runs into a problem when there are \emph{two subsequent charges with the same sign.} 
To generalize the mapping to these cases, we introduce extra \emph{defect} degrees of freedom on the sites,
which keep track of those charges that do not conform to the rule of alternating signs. 
To do this, we sweep through the chain from left to right, putting spins on the bonds in accordance with the previous rule.
When, at some position $j$, we encounter a charge that has the same sign as the one preceding it, we fix the spin of the bond $(j,j+1)$ to coincide with preceding one, $\sigma_{j,j+1}^z = \sigma_{j-1,j}^z$.
At the same time, in order to keep track of the charge, we place a \emph{defect} on the site $j$. 
This way we end up with a model with two types of degrees of freedom: spins on the bonds and defects on the sites. 
The resulting Hilbert space is $3^L$ dimensional\footnote{There is some ambiguity regarding the completely empty state: by convention we choose it to correspond to a state with all bond spins pointing right and no defects.}, since a site combined with the bond on its right only have together three possible configurations.
An example of this mapping with four defects is shown in Fig.~\ref{fig:mapping_sketch}(b).
 
It is important to note that while defects themselves do not carry a sign, we can still distinguish whether they correspond to positive or negative charges in the original variables by looking at the spins surrounding them: a defect with neighboring spins pointing right is mapped to a positive charge, while a defect with neighboring spins pointing left is mapped to a negative charge.
We refer to these as $(+)$- and $(-)$-defects, and they correspond to eigenvalues $\pm 1$ of the operator $n^d_j \sigma^z_{j,j+1} = \sigma^z_{j-1,j} n^d_j$. 
The old and new degrees of freedom are related to each other by the generalized Gauss's law
\begin{equation}
\label{eq:GGL} \frac{1}{2}\big(\sigma_{j,j+1}^z - \sigma_{j-1,j}^z\big)  = S_j^z - \sigma^z_{j-1,j} n^d_j,
\end{equation} 
 which allows us to write the global charge and dipole moment in terms of the new variables as
\begin{align} \label{eq:Qmap}
Q&=\frac{1}{2}\big(\sigma^z_{L,L+1} - \sigma^z_{0,1} \big) +\sum_{j=1}^L \sigma^z_{j-1,j} n^d_j,\\ \label{eq:Pmap} P&=-\frac{1}{2}\sum_{j=0}^{L-1}\big(\sigma_{j,j+1}^z-\sigma^z_{L,L+1}\big) + \sum_{j=1}^L j\sigma^z_{j-1,j} n^d_j.
\end{align}
Notice that in the absence of defects, $Q$ is set entirely by the configuration of the bond spins on the boundaries, while $P$ maps onto the total magnetization (up to a constant), i.e., a usual global U(1) internal symmetry.

The mapping we defined is clearly a non-local one. A natural question to ask is: when is the resulting Hamiltonian local in the new variables? In fact, the relevant property of $H$ that ensures this is the same as the one encountered above as a necessary condition for statistically localized strong boundary modes. Namely, we require the following condition: terms of the Hamiltonian acting on a given region of space can not change the sign of the left- and rightmost charges within this region. Indeed, it was already noted in Ref. \onlinecite{Sala19} that $H_3$ satisfies this property. Consequently, $H_3$ also conserves $\hat{q}_{\ell,r}$ and therefore exhibits strong boundary modes. We return to this point below.

\subsection{Labeling of connected sectors}\label{sec:H3labels}

Armed with this mapping, we can now identify the integrals of motion that label the fragmented Hilbert space, and show how they fit into the general notion of statistically localized operators discussed above.

\subsubsection{Pattern of defects}

We start by noting that the Hamiltonian in Eq.~\eqref{eq:H3} does not contain any terms that could create or destroy defects: \emph{the number of defects, $N^d\equiv\sum_j n_j^d$, is conserved}. 
This can be confirmed explicitly by considering the effect of local terms in $H_3$.
Thus the number of defects acts as an emergent U(1) symmetry (different from the original U(1) symmetry of charge conservation), emergent in the sense that it is non-local in the original variables and only becomes local after the mapping outlined above. One can use the operators $\hat{q}_k$, defined for the physical variables in Eq.~\eqref{eq:qk_def}, to express the number of defects as
\begin{equation} \label{eq:Nd} N^d=\frac{1}{2}\sum_{k=1}^L  \big(\hat q_{k+1}\big)^2\big(1+\hat q_k \hat q_{k+1} \big).
\end{equation}
This further emphasizes the non-local nature of the defects.

In fact, the Hamiltonian $H_3$ conserves not only the total number of defects, but also the pattern of their signs (similarly to how $H_{t-J_z}$ conserved not just the number of fermions, but also the spin orientation of each fermion). For example, the state shown in Fig.~\ref{fig:mapping_sketch}(b), with (from left to right) a $(-)$-defect followed by three $(+)$-defects, can only go to configurations with the same pattern. Thus we see that the mechanism behind the fragmented Hilbert space is analogous in the two cases, except that for $H_3$ it originates from a `hidden', rather than explicit, $U(1)$ symmetry. 

The pattern of defects can be characterized by eigenvalues of statistically localized operators, similar to the ones discussed above in the case of the $t-J_z$ model. 
In fact, after mapping to bond spins and defects, one can directly use the same set of operators to label the defect patterns, as defined in Eq.~\eqref{eq:qk_def}, by replacing $S_j^z$ with the local defect charge operator $\sigma_{j-1,j}^z n_j^d$ and $\hat{\mathcal{P}}^k_j$ with a projector onto configurations with $\sum_{i<j} n_i^d = k-1$ and $n_j^d=1$.
In the original variables, these are rather complicated non-local operators. 
Nevertheless, a Haar random state in the thermodynamic limit will have a finite density of defects, $\nu_d \equiv \braket{N^d} / L = \frac{1}{3}$ (see App.~\ref{app:Haar}). 
Indeed, since for large $L$ the variance is once again exponentially suppressed ($\mathbb{E}_\text{Haar}[\braket{n_j^d}^2] - \mathbb{E}_\text{Haar}[\braket{n_j^d}]^2 \propto 3^{-L}$), almost all states have a similar defect density.
For such states, one could repeat the argument in Sec.~\ref{sec:slim} to argue that the probability distribution of finding the $k$-th defect on site $j$ is peaked around a position $j = k/\nu_d$, with a width that scales as $\sqrt{k}$.
Similarly, a random state with a fixed total charge $Q$ will also have a finite $\nu_d$ and therefore leads to a partially localized probability distribution. 
Thus the operators that label the defect patterns and the corresponding Hilbert space sectors of $H_3$ are statistically localized in the sense we defined previously.

We conclude this section by noting that apart from the charges of each defect, $H_3$ also conserves the sign of the leftmost and rightmost \emph{physical} charges, as measured by the operator $\hat{q}_\ell$ and $\hat{q}_r$ defined in Eqs.~\eqref{eq:ql_def} and~\eqref{eq:qr_def} respectively (as mentioned above, this condition is in fact necessary to ensure that the Hamiltonian remains local after mapping to the new variables). 
This implies that our conclusions about the lack of thermalization at the boundary, and about exact degeneracies in the spectrum, discussed in Sec.~\ref{sec:zero_modes} for the $t-J_z$ model, apply also to $H_3$. 
However, $H_3$ is different from $H_{t-J_z}$, in that it shows fully localized behavior also \emph{in the bulk}.
To understand the reason for this, we now turn to a further set of conserved quantities possessed by $H_3$.

\subsubsection{Dipole moment of dynamical disconnected regions}\label{sec:local_dipoles}

While the conservation of the pattern of defect charges is sufficient to fragment the Hilbert space into exponentially many disconnected sectors, it does not account for all the sectors of $H_3$.
The conservation of the signs of defects (which are in fact a subset of the conserved quantities exhibited by $H_{t-J_z}$) is also insufficient to explain the localized behavior (i.e., infinitely long-lived autocorrelations) occuring in the \emph{bulk}, which was observed previously~\cite{Sala19}.
As we now argue, this rich non-ergodic dynamics originates from an interplay between the SLIOMs discussed in the previous section (that is, the pattern of defects), and the conservation of the total dipole moment.
Thus, while on their own neither of those ingredients leads to fully localized behavior, their combination is sufficient to make $H_3$ localized.

The fact that dipole conservation leads to further disconnected sectors can already be seen in the case of states with no defects, $N^d=0$.
As seen from Eqs.~\eqref{eq:Qmap} and~\eqref{eq:Pmap}, the zero defect sector with a given boundary condition (and thus fixed total charge $Q=0,\pm 1$) further splits up into sectors according to the total magnetization of the bond spins, $\sum_j \sigma_{j,j+1}^z$, which in this case is equal to the dipole moment $P$ up to a constant shift. 

 \begin{figure}[t!]
	\centering
	\includegraphics[width = 1.\columnwidth]{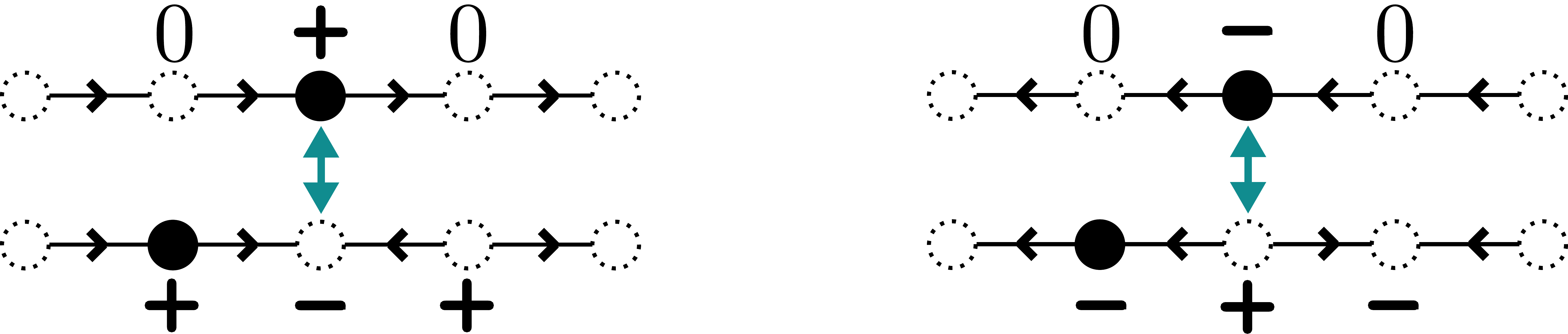}
	\caption{\textbf{Hopping of defects.} To maintain the constraints, when a defect hops it has to flip a bond spin to its right, making its dynamics asymmetrical. In the original variables, this process is equivalent to emitting/absorbing a dipole \emph{from the right}.}
	\label{fig:hopping_sketch}
\end{figure}

When defects are present, they also carry a dipole moment, as shown by Eq.~\eqref{eq:Pmap}.
Dipole conservation then puts further constraints on the ways in which defects are allowed to move in the system: whenever a defect hops to a neighboring site, this has to be accompanied by a spin flip, in order to ensure that the overall dipole is conserved, e.g. $\ket{\rightarrow \rightarrow \bullet \rightarrow} \Leftrightarrow \ket{\rightarrow \bullet \rightarrow  \leftarrow}$.
This corresponds to the fact that in the original variables, charges can only hop by emitting dipoles, as illustrated in Fig.~\ref{fig:hopping_sketch}. 
However, due to the asymmetric definition of the defect---same charge as the nearest \emph{on its left}---its hopping only modifies the configuration on bonds that are to its right.
This is the same as saying that defects can only emit (absorb) dipoles to (from) their right and never from their left.
Thus, for every defect \emph{the total dipole moment of charges to its right (including the defect itself) is conserved}.
This implies that the dipole to the left of the defect (\emph{not} counting the defect) is also separately conserved.

\begin{figure}
	\includegraphics[width = 1\columnwidth]{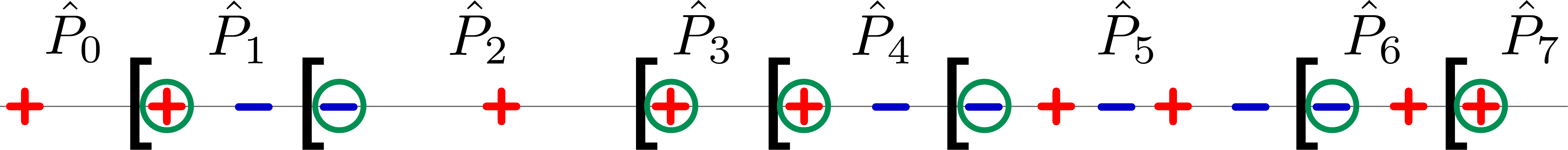}
	\caption{\textbf{Labeling of connected sectors in the original variables.} Charges that have the same sign as the ones to their left (circled) correspond to \emph{defects}, whose total number ($N^d$) and pattern is conserved by the Hamiltonian $H_3$. Moreover, the dipole moment $\hat P_k$ within each region between two subsequent defects (including the defect on the left but not the one on the right, as indicated by the brackets) is also independently conserved, such that the total dipole becomes $P=\sum_{k=0}^{N^d}\hat P_k$.}
	\label{fig:fig6}
\end{figure}

We thus find that each defect gives rise to an additional conserved quantity.
Equivalently, we could take a configuration with $N^d$ defects, which separate the chain into $N^d+1$ regions, and associate a conserved dipole moment to each of these regions. 
In assigning the dipole moment $\hat P_k$ to the region between defects $k$ and $k+1$, one should include the $k$-th defect (at the left boundary) but not the $(k+1)$-th on its right (e.g. $\ket{\cdots\bm{[}\bullet \rightarrow \cdots \leftarrow \bm{)}\bm{[}\bullet \leftarrow \cdots}$).
The total dipole moment then becomes\footnote{By definition, $\hat P_0$ corresponds to the dipole moment between the left boundary of the chain and the first defect; while $\hat P_{N^d}$ corresponds to the dipole moment between the last defect and the right boundary.} $P=\sum_{k=0}^{N^d}\hat P_k$, where $k$ labels a region separated by defects, each with its own conserved dipole moment $\hat P_k$\footnote{Note that, while the total charge $\hat Q_k$ in each region is also conserved, this does not give rise to new independent constants of motion, since the value of these charges are already fixed by the pattern of defects.}.
This is shown in Fig.~\ref{fig:fig6} in terms of the original spin-1 degrees of freedom.

Note that, while the position of the $k$-th defect in the bulk has fluctuations that grow with system size as $\propto \sqrt{L}$ (much like the case of the $k$-th charge in the $t-J_z$ model before), the average \emph{distance} between neighboring defects remains finite in the thermodynamic limit for states with a finite defect density $\nu_d$. 
We can make this point more explicit, by defining the operator that measures $\hat P_k$ as
\begin{equation}\label{eq:Pa_def}
\hat{P}_k = \sum_{ij} \hat{\mathcal{Q}}^k_{ij} P_{ij},
\end{equation}
where $\hat{\mathcal{Q}}_{ij}^k$ is a projector onto configurations where the $k$-th defect sits on site $i$ and the $(k+1)$-th defect is on site $j$, while $P_{ij}$ measures the dipole moment in the region $[i,j-1]$ (including the former but not the latter defect).
Given Eq.~\eqref{eq:Pa_def}, we can go to center of mass and relative coordinates: while the expectation value $\braket{\hat{\mathcal{Q}}_{ij}^k}$, as a probability distribution, is only partially localized in $\frac{i+j}{2}$, it is \emph{exponentially} localized in the relative coordinate, decaying as $(1-\nu_d)^{-(j-i)}$.
In this sense, $\hat{P}_k$ is statistically localized to \emph{a finite region} (see App.~\ref{app:slim_def} for more details on the definition of SLIOMs appropriate to this case).
As we show in the next section, he existence of these additional conserved quantities additional dynamical constraints on the mobility of defect configurations. These constraints, together with the statistical localization of $\hat{P}_k$, account for the fact that $H_3$ has infinite coherence times for charge autocorrelations in the bulk, (as well as a broad distribution of entanglement in energy eigenstates, which we discuss in Sec.~\ref{sec:types}), as  previously observed in Refs.~\onlinecite{Sala19, Vedika19}.

To summarize, let us compare the conservation laws of $H_3$ with those of the $t-J_z$ model discussed above. 
In the latter case, we had a conserved number of fermions, each of which carries a spin-1/2 whose $S^z$ components are all separately conserved --- defining what we have named the \emph{pattern of spins}.
$H_3$ is different for two reasons. 
First, the objects, whose pattern is conserved are the defects, which are non-local in the original variables. 
Furthermore, $H_3$ has an additional set of conserved quantities $ \{\hat P_k\}_{k=0}^{N^d}$, arising due to the interplay between dipole conservation and the defect pattern: all the spatial regions separated by defects have separately conserved dipole moments.
Altogether, we have identified the following set of conserved quantities for $H_3$: the total charge $Q$ and dipole $P$, the left- and rightmost charges $\hat{q}_{\ell,r}$, the number of defects $N^d$, the charge of each defect $\{Q_k=\pm 1\}_{k=1}^{N^d}$ and the dipole moment of regions between defects, $\{\hat P_k\}_{k=0}^{N^d}$. 
We have numerically confirmed that these integrals of motion together \emph{uniquely} label \emph{all} the connected sectors of $H_3$ in the local $S^z$ basis.
Since a dipole-conserving random circuit of 3-site gates has the same~\cite{Sala19} fragmentation of the Hilbert space as $H_3$, it consequently also conserves all of the quantities identified above.
%

\subsection{Implications for dynamics} \label{sec:types}

In the previous section we saw how the conserved quantities of $H_3$ fit into the scheme of SLIOMs (see also App.~\ref{app:slim_def}). However, their precise nature is different from the simpler case of the $t-J_z$ model discussed in Sec.~\ref{sec:tJz}. As mentioned above, this difference is responsible for the fact that, despite both being strongly fragmented, the two models exhibit rather different dynamics in their bulk: $H_3$ has infinite correlation times~\cite{Sala19,Vedika19}, unlike $H_{t-J_z}$. Here we explain how the SLIOMs constructed in the previous section bring about localized dynamics, highlighting the role played by the dipole moments $\hat P_k$.

\subsubsection{Charge localization}

To see how the conservation laws lead to localized behavior, consider a configuration where there are two subsequent defects with a $+$ charge, at sites $i$ and $j>i$. By the definition of defects, the region $[i+1,j-1]$ between them has $0$ total charge and a dipole moment $p\geq 0$. As long as the position $i$ is fixed, $p$ is conserved. This dipole cannot be compressed to a region of less than $p$ sites, forcing the position of the second defect to obey $j > i + p$. But the right hand side of this inequality is in fact one of the conserved quantities $\hat P_k$, and therefore time-independent\footnote{Note that the condition of having zero total charge in the middle region is important, as it allows us to always shift the reference frame and measure $p$ from the position $i$.}. Therefore, the position $j$ of the second defect can \emph{never} cross this particular location and remains restricted to half of the chain. Similarly, since $p\geq 0$ at all times, and $i+p$ is conserved, we have that the left defect can move at most $p$ sites to the right. Clearly, the same argument applies to a pair of $(-)$-defects\footnote{For two defects with \emph{opposite} signs, one gets a weaker constraint $j-i > P_k+1$, i.e., a lower bound on their \emph{distance}.}. 

Let us now consider a defect somewhere in the bulk of the chain for a typical configuration in the $z$-basis. How far can it travel to the left? If the nearest defect to its left is of the same sign, it constrains its motion by the above argument. More generally, consider the closest pair of subsequent equal sign defects on the left; due to the hard-core constraint, these restrict the motion of all defects to their right, including the original one. Therefore, the only way for a given defect to travel a distance $\ell$ to the left is if all the defects originally within this region have an exactly alternating sign pattern. However, the relative number of such configurations scales as $e^{-\gamma \ell}$ for some constant $0 < \gamma < 1$, and therefore, with probability $1$ in the thermodynamic limit, $\ell$ cannot be larger than $O(1)$. The same argument applies to travelling to the right, which shows that almost all defects are localized to finite regions\footnote{Note that one could also define defects starting from the right, rather than the left, edge of the chain. These could be used to further constrain the possible transitions.}.

Consider now the infinite temperature charge autocorrelator. We can expand it in terms of product states $\ket{\mathbf{s}} = \bigotimes_i \ket{s_i}$ in the original variables (i.e., $s_i = +,-,0)$ as
\begin{multline}
    \braket{S_j^z(t) S_j^z}_{\beta = 0} \\= \frac{1}{3^L} \left[ \sum_{\substack{\mathbf{s}\\ s_j = +}} \braket{\mathbf{s}(t)|S_j^z|\mathbf{s}(t)} - \sum_{\substack{\mathbf{s}\\ s_j = -}} \braket{\mathbf{s}(t)|S_j^z|\mathbf{s}(t)} \right].
\end{multline}
In half of the cases, the initial $+$ charge on site $j$ is a defect. In that case, as the above argument shows, it is almost surely restricted to live in a final spatial region with an overall charge of $+1$, thus yielding a positive contribution to the autocorrelator. If the size of the region is $\ell$, the contribution is expected to be $O(1/\ell)$, and in the thermodynamic limit, their sum gives $\sum_{\ell=1}^\infty e^{-\gamma\ell}/\ell = -\ln(1-e^{-\gamma}) > 0$. There is another equal contribution stemming from the $(-)$-defects. This shows that the SLIOMs lead to charge localization even at infinite temperature\footnote{One could attempt to derive the same result by applying Mazur's inequality, using all the diagonal conserved quantities of $H_3$.}.

\subsubsection{Entanglement growth}

\begin{figure}
	\centering
	\includegraphics[width=0.99\linewidth]{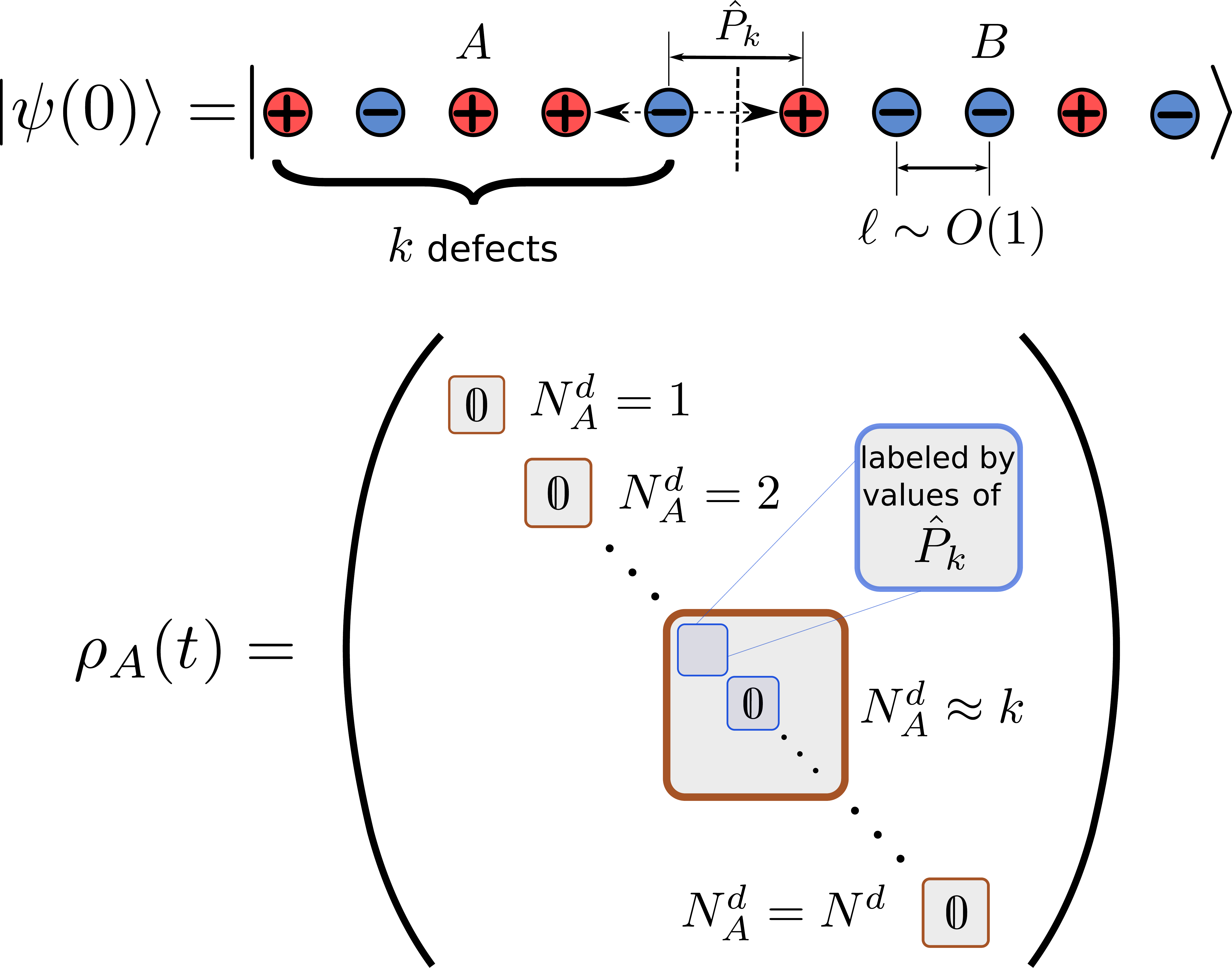}
	\caption{\textbf{Entanglement growth for $H_3$.} The saturation value of the half-chain entanglement at long times for the dipole-conserving Hamiltonian $H_3$ (Eq.~\eqref{eq:H3}) for initial product states in the $S^z$ basis can be understood from the emergent conservation of the number of defects $N^d$, along with the SLIOMs $\hat P_k$ introduced in Sec.~\ref{sec:local_dipoles}. The former implies a block-diagonal structure of the reduced density matrix $\rho_A$ on region $A$ (chosen to be half the chain), of the form $\rho_A=\oplus_{N^d_A=0}^{N^d}\rho_A(N_A^d)$. However, due to the kinetic constraints on the mobility of defects (see main text), only a few of these blocks are non-vanishing, those where $N^d_A$ is close to its value in the initial state. The additional conservation of dipole moment within a region between defects $\{ \hat{P}_k\}$ further block diagonalizes $\rho_A(N_A^d)$, most of which are again zero.}
	\label{fig:eeexplanation}
\end{figure}

Another signature of localized behavior in $H_3$ is the numerical observation~\cite{Vedika19,Sala19} that the entanglement entropy of the long-time steady state is sub-thermal, even for an initial random product state that is not in the $z$-basis and therefore has weight in all the connected sectors. In Ref.~\onlinecite{Vedika19} it was argued that this saturation value is determined by the size of the largest sector, and therefore should scale as $\frac{L}{2}\ln(2)$ for $H_3$, which is consistent with the numerical results (see App.~\ref{app:ScaEE}). 

However, the block structure of the Hamiltonian itself does not put any constraints of the amount of entanglement it can generate. 
In particular, even a unitary made up entirely by random diagonal phases in the $z$-basis can generate the same amount of entanglement as a Haar random unitary, when applied to a state that is an equal weight superposition of all basis states\footnote{TR thanks Andr\'as Gily\'en for a very useful discussion on this topic.}~\cite{TomasiMBL}. 
This point is also illustrated by considering the $t-J_z$ model. 
In that case, even though the dimension of the largest connected component is only $2^L$, for an initial (Haar) random product state, the von Neumann entropy saturates to a value much larger than $\frac{L}{2}\ln(2)$, as we show in App.~\ref{app:ScaEE}.

These examples show that, in order to explain the sub-thermal entropy exhibited by $H_3$, one has to combine the knowledge of the conserved quantities with considerations of spatial locality. 
Indeed, going back to the completely diagonal case, if we restrict ourselves to local terms of range at most $\ell$, the amount of entropy they can produce is upper bounded by $(\ell-1)\log{d}$ (where $d$ is the on-site Hilbert space dimension). 
In a similar manner, it appears that combining all the conservation laws of $H_3$ with the restriction of spatial locality is sufficient to prevent the state from reaching maximal entropy density.
Since we saw that the conserved dipole moments $\hat P_k$ are largely responsible for the localization of the charge degrees of freedom, it is expected that they are responsible for constraining entanglement growth. 

The fact that the conservation laws severely restrict entanglement can be easily seen in the case of evolving from an initial product state \emph{in the $z$-basis} with $H_3$. Such a state has a well defined quantum number for all SLIOMs. Consequently, the reduced density matrix of a bi-partition can be block diagonalized by e.g., the number of defects on one side. As noted in Sec.~\ref{sec:H3labels}, for a randomly chosen $z$-product state, which has a finite density of defects, the movement of almost all defects will be restricted to $O(1)$ regions due to the conservation laws. Therefore, a particular entanglement cut can only be crossed by a small subset of defects, and consequently many of its blocks, will be identically zero. Furthermore, each block with $k$ defects to the left of the cut can be further decomposed into smaller blocks using the conserved dipole moment $\hat P_k$ (see Fig.~\ref{fig:eeexplanation}). Since the $k$-th defect can only travel a finite distance to the left, it can only emit a finite number of dipoles, such that the reduced density matrix for most initial configurations is restricted to a few blocks of size $O(1)$. Consequently, it only has a finite number of non-vanishing eigenvalues, limiting its entanglement to an area law. The same argument explains the broad distribution of entanglement entropies observed for the eigenstates of $H_3$~\cite{Sala19,Vedika19}.

The above discussion shows that the structure of SLIOMs we uncovered gives serious restrictions for entanglement growth for initial states in the $z$ bases. 
We expect the same mechanism to be responsible also for the sub- thermal saturation value for completely random product states.

\subsection{Largest sectors and SPT order}\label{sec:spt}

A particular corollary of the discussion in Sec.~\ref{sec:H3labels} is that increasing the number of defects decreases the connectivity of the Hilbert space, since each new defect leads to a further conservation law (the associated dipole moment), which one needs to fix in order to specify a sector. 
Indeed, one can check numerically that the largest connected sectors all have zero defects. 
Moreover, we confirm numerically that the overall ground state of $H_3$ (which is 4-fold degenerate, as we discuss below) also belongs to these four largest sectors. 
Motivated by this, we now turn our attention to the subspace with no defects. 

In fact, $H_3$ takes a particularly simple form within this subspace. 
Since there are no defects, the only degrees of freedom are the bond spin-1/2's, which can take any configuration. 
As one can check by considering each local term, $H_3$ simply becomes
\begin{equation} \label{H3_rest}
\left. H_3\right|_{N^d=0} = - 2\sum_{j=2}^{L-1} \left( \sigma_{j-1,j}^x \sigma_{j,j+1}^x + \sigma_{j-1,j}^y \sigma_{j,j+1}^y \right),
\end{equation}
i.e. a spin-$1/2$ XY model on a chain of length $L-1$ (note that the two auxiliary spins, $\sigma^z_{0,1}$ and $\sigma^z_{L,L+1}$ do not appear in the Hamiltonian), exactly solvable via a Jordan-Wigner transformation to free fermions\footnote{One could the same mapping for the $t-J_z$ Hamiltonian; in particular, for $J_z = 0$ one finds that $\left. H_{t-J_z}\right|_{N^d=0} \propto \sum_{j=1}^{L-1} \left( \sigma^x_{j,j+1} - \sigma^z_{j-1,j} \sigma^x_{j,j+1} \sigma^z_{j+1,j+2}\right)$. This Hamiltonian describes a critical point between the 1D cluster phase and a trivial paramagnet, which is another way of seeing that $H_{t-J_z}$ is gapless (as one can confirm numerically, its ground state is indeed in the $N^d = 0$ sector).}.
This Hamiltonian conserves $Z_\text{tot} \equiv \sum_j \sigma_{j,j+1}^z$, equal to the dipole moment in the original model, with the largest symmetry sector being the one with half-filling ($Z_\text{tot}=0$)\footnote{The dimension of the largest connected sector is therefore (assuming an odd number of sites) $\binom{L-1}{(L-1)/2}$, scaling asymptotically as $\propto 2^L$ up to logarithmic corrections. This confirms earlier numerical results~\cite{Vedika19,Sala19}.}. 
The ground state of this model is gapless due to the presence of Fermi points and has an effective low energy Luttinger liquid description.
We confirm that this is also the ground state of $H_3$ overall, by finding the ground state in DMRG and comparing its energy with that of the ground state of the XY chain at half filling, finding perfect agreement.

However, this is not the full story. As mentioned above, the ground state has a 4-fold degeneracy. In fact, this is true for all eigenstates within the zero defect sector: as seen above, this sector consists of 4 equivalent XY chains with 4 different boundary conditions. These corresponds to the four possible choices of the leftmost and rightmost charge in the system, which are conserved under $H_3$. Moreover, we find numerically that even eigenstates \emph{with} defects are 4-fold degenerate throughout the entire spectrum. This degeneracy is due to zero modes at the boundaries of an open chain, and is not present with periodic boundary conditions\footnote{$H_3$ still has a significant amount of degeneracies with PBC, but it also has non-degenerate eigenvalues.}. Nevertheless, the exact 4-fold degeneracy is specific to $H_3$ and can be lifted to a 2-fold degeneracy by adding perturbations, diagonal in the $S^z$-basis, which preserve the block structure of $H_3$. The 2-fold degeneracy, on the other hand, is robust as long as we preserve the spin rotation symmetry $R_x = e^{i\pi\sum_j S_j^x}$ and the signs of the left- and rightmost charges, analogously to the case of the $t-J_z$ model discussed before.

The strong zero modes at the boundary appear concurrently with symmetry protected topological (SPT) order in the bulk, for \emph{all} eigenstates inside the no defect subspace. This can be seen by considering the string order parameter, $\braket{S_i^z e^{i\pi\sum_{l=i+1}^{j-1} S_l^z}S_j^z}$. This measures the `hidden antiferromagnetic order' of the Haldane phase, which becomes apparent after dropping all the empty sites. States with no defects have such a hidden AFM order by construction. More formally, acting on states without defects, the string factorizes due to the Gauss's law~\eqref{eq:GGL} as $e^{i\pi\sum_{l=i+1}^{j-1} S_l^z} \to \sigma_{i,i+1}^z \, \sigma_{j-1,j}^z$, an explicit example of symmetry fractionalization. Consequently, the string order parameter simplifies to $\braket{(1-\sigma^z _{i-1,i} \sigma^z _{i,i+1}) (\sigma^z _{j-1,j} \sigma^z _{j,j+1}-1)}/4$. In the limit $|j-i| \to \infty$ this factorizes into the product of local expectation values. Now, the expectation value $\braket{1-\sigma^z _{i-1,i} \sigma^z _{i,i+1}}$ is non-zero for any translation invariant state, except for a completely spin polarized one (i.e. the empty state in the original variables). Therefore, all eigenstates with $N^d=0$, except for the completely empty state, have (symmetry protected) topological order\footnote{In principle the non-vanishing string order parameter is also compatible with the symmetry being spontaneously broken. However, in our case, within the zero defects sector the symmetry acts trivially in the bulk and thus we associate the presence of string order with a symmetry protected topological state.}. This is reminiscent to the appearance of topological order in excited states of MBL systems~\cite{Huse2013,Chandran2014}.

Relatedly, the ground state of $H_3$ is a \emph{gapless topological phase}~\cite{Verresen19}, similarly to the case of $H_{t-J_z}$ discussed before. The separation of degrees of freedom into bond spins and defects provides a simple interpretation of this: while the former are gapless, the latter are gapped and are responsible for protecting the SPT order in the ground state. This latter fact can be seen by noting that the symmetry $R^z \equiv e^{i\pi\sum_i S_i^z} $ of the Hamiltonian becomes (in the full Hilbert space, including defects) $R^z = \sigma_{0,1}^z e^{i\pi \sum_i n_i^d} \sigma_{L,L+1}^z$. This is therefore a gapped symmetry in the nomenclature of Ref.~\onlinecite{Verresen19}, in the sense that operators charged under this symmetry in the bulk necessarily create gapped excitations (in this case, defects). The coexistence of gapless bulk with these additional gapped degrees of freedom ensures the two-fold degeneracy of the ground state, up to an exponentially small finite size splitting~\cite{Vasseur17,Verresen19}. In this particular model, due to the boundary SLIOMs, this degeneracy is \emph{exact} (and present throughout the spectrum). Perturbations, which destroy the SLIOMs but preserve the $\mathbb{Z}_2 \times \mathbb{Z}_2$ symmetry of $\pi$-rotations will keep the twofold degeneracy at low energies, now exhibiting the aforementioned exponentially small finite-size splitting.

\section{Summary and outlook}\label{sec:outlook}

In this work, we explicitly constructed integrals of motion for two models that exhibit the phenomenon of strong Hilbert space fragmentation, including a complete description of the Hamiltonian $H_3$ introduced in Ref.~\onlinecite{Sala19}. 
These integrals of motion label the different disconnected sectors of the many-body Hilbert space, playing a role analogous to local integrals of motion in many-body localized systems. They are dominated by contributions from a sub-extensive region in space, but in such a way that the location and width of this region can be tuned by, for example, changing the average filling fraction in the system. This lead us to term these observables \emph{statistically localized}.

These statistically localized integrals of motion (SLIOMs) lead to a breakdown of eigenstate thermalization in both models we study. However, their effect on autocorrelations in the bulk depends on the nature of their distribution, which leads to different behavior for the two models. In the $t-J_z$ model (which we argued can be realized in Rydberg atom experiments), all SLIOMs in the bulk are localized to regions of size $O(\sqrt{L})$. As a result, autocorrelations saturate to values $O(1/\sqrt{L})$, which are anomalously large compared to generic thermalizing systems, but nevertheless vanish as $L\to\infty$. For the dipole-conserving Hamiltonian $H_3$, on the other hand, some of the bulk conserved quantities are effectively localized to $O(1)$ regions and lead to finite autocorrelations even in the thermodynamic limit.

SLIOMs near the boundary, on the other hand, are localized to finite regions and lead to infinitely long coherence times for \emph{both} models. We showed that these boundary SLIOMs can survive certain perturbations that destroy the strong fragmentation in the bulk, defining a statistically localized analogue of strong zero modes, where a thermalizing bulk co-exists with an explicitly non-ergodic boundary. We also analyzed the relationship between these zero modes and the ground states of the two models, which exhibit symmetry protected topological order, despite being gapless. 

Several questions remain to be explored. Dipole-conserving spin-1/2 chains with 4-site terms show similar behavior as $H_3$, and therefore one can expect that it is possible to construct analogous SLIOMs in that case. On the other hand, it is unclear whether the scheme presented here could be used to find the conserved quantities relevant for longer-range generalizations of $H_3$ (which exhibit weak fragmentation~\cite{Sala19}). Even within the subset of strongly fragmented models (i.e., with the largest symmetry sector being a vanishing fraction of the full Hilbert space), qualitatively very different behaviors can arise, as the two examples in our paper demonstrate. Therefore, it would be interesting to develop a more quantitative understanding of different `degrees' of fragmentation, as these have clear effects on the spreading of correlations. The structure of conservation laws we uncovered could also be useful for understanding the dynamics of entanglement and operator growth in these systems.

Another direction is to explore the stability of the boundary SLIOMs to additional perturbations, i.e., whether they can still lead to unusually long coherence times even when they are not explicitly conserved. More generally, it would be interesting to investigate the role SLIOMs play in a many body localized phase~\cite{Iadecola2018, Tomasi19}, both at the boundary and in the bulk. In fact, our construction of SLIOMs for the $t-J_z$ model also applies to the fragmented Hilbert space studied in Ref.~\cite{Tomasi19}, after mapping the onsite fermionic to spin degrees of freedom for open boundary conditions. In fact, we expect that this construction holds for certain strong-coupling expansions of 1D Hamiltonians. It would also be interesting to look for other models exhibiting SLIOMs, either at their boundary or in their bulk. 

\acknowledgements 

The authors thank  Giuseppe De Tomasi, Christopher J. Turner, Johannes Feldmeier, Andr\'as Gily\'en, Marcos Rigol and Johannes Zeiher for discussions.
We also thank Mari Carmen Bañuls, Maksym Serbyn and Marko \v{Z}nidari\v{c} for useful comments on the manuscript. 
We acknowledge support from “la Caixa” Foundation (ID 100010434) fellowship grant for post-graduate studies (P.S.), the Harvard Quantum Initiative Postdoctoral Fellowship in Science and Engineering (RV), the Technical University of Munich - Institute for Advanced Study, funded by the German Excellence Initiative, the European Union FP7 under grant agreement 291763, the Deutsche Forschungsgemeinschaft (DFG, German Research Foundation) under Germany's Excellence Strategy--EXC-2111--390814868, Research Unit FOR 1807 through grants no. PO 1370/2-1, TRR80 and DFG grant No. KN1254/1-1, and DFG TRR80 (Project F8), and from the European Research Council (ERC) under the European Union’s Horizon 2020 research and innovation programme (grant agreements No 771537 and 851161).
This research was conducted in part at the KITP, which is supported by NSF Grant No. NSF PHY-1748958

\twocolumngrid

\appendix

\section{More refined definition of SLIOMs}\label{app:slim_def}

While in Sec.~\ref{sec:slim} we gave a definition of SLIOMs, sufficient for the $t-J_z$ Hamiltonian, it is worthwhile to elaborate further on the structure of the SLIOMs we encountered in this work and how precisely localization appears for them. 

In the case discussed in Sec.~\ref{sec:slim}, a very useful property was that the terms appearing in the definitions of the SLIOMs $\hat{q}_k$ squared to projectors ${\mathcal{O}^k_i}^\dagger \mathcal{O}^k_i = \hat{\mathcal{P}}^k_i$ (using the convention in Eq.~\eqref{eq:SpinOp}). These projectors were then used to define the spatial distribution over $i$ that we analyzed in the main text. However, one could consider a slightly more general version of the $t-J_z$ model, where the fermions carry a higher spin, $S > 1/2$. In that case, $(S_i^z)^2$ is no longer equal to the projector $\tilde{n}_i$, and the interpretation becomes less clear.

In this more general case, we can still use the definition of $\hat{q}_k$ introduced in the main text:
\begin{equation}
\hat{q}_k = \sum_i \hat{\mathcal{P}}^k_i S_i^z.
\end{equation}
Note that the conserved quantity splits up into a projector ($\hat{\mathcal{P}}^k_i$) onto certain configurations and an associated `charge' ($S_i^z$), and that in our discussion of the statistical localization it was in fact only the projector part that played a role. Note that this is analogous to the structure we observed for the local dipole moments defined for the Hamiltonian $H_3$ in Eq.~\eqref{eq:Pa_def}, i.e., a sum of projectors multiplied by an associated `charge' (in that case, the dipole moment between two subsequent defects). In both cases, the statistical localization is a property of the projectors, rather than the charges.

This suggests the following general definition of SLIOMs that encompasses all the cases encountered in our manuscript:
\begin{equation}\label{eq:slim_general_def}
\hat{q} = \sum_{i_1,i_2,\ldots,i_n} \hat{\mathcal{Q}}_{i_1i_2\ldots i_n} C_{i_1i_2\ldots i_n}.
\end{equation}
Here, $\hat{\mathcal{Q}}_{i_1i_2\ldots i_n}$ is a projection onto configurations where the sites $i_1,\ldots,i_n$ are occupied by a particular combination of particles, while $C_{i_1i_2\ldots i_n}$ is some charge (in the cases we consider, usually an integer) associated to this configuration. One can then consider the distribution of the expectation value (in some appropriately chosen ensemble of typical states) $\braket{\hat{\mathcal{Q}}_{i_1i_2\ldots i_n}}$. This is now a distribution on $[1,L]^n$ (where we have assumed a 1D system) and one can examine how it is localized on this potentially larger dimensional space.

The $t-J_z$ model with arbitrary spin corresponds (for a given fermion indexed by $k$) to the choice $n=1$, $\hat{\mathcal{Q}}_i = \hat{\mathcal{P}}^k_i$ and $C_i = S_i^z$. The sign of a defect in $H_3$ again corresponds to taking $n=1$, but now with $\hat{\mathcal{Q}}_i$ a projector onto having the $k$-th defect on site $i$ and again $C_i = S_i^z$ (in the original spin-1 language). Both of these cases are partially localized, to regions of size $O(\sqrt{L})$, in the single coordinate $i$. The localized dipole moments, on the other hand, correspond to $n=2$, with $\hat{\mathcal{Q}}_{ij}$ projecting onto configurations with a pair of defects on sites $i,j$ with no other defect in-between. The expectation value $\braket{\hat{Q}_{ij}}$ in this case is exponentially localized in the relative coordinate $j-i$ as discussed in the main text. The associated charge is now the dipole moment $C_{ij} = P_{ij} \equiv \sum_{\ell=i}^{j-1} \ell S_\ell^z$. 

This general definition also allows us to talk about conserved quantities for the $t-J_z$ model with periodic boundaries (see also App.~\ref{app:pbc}), even though in this case they are no longer localized. With periodic boundaries, we no longer have a way of labeling fermions individually (e.g. the first fermion can become the last by travelling around the boundary). Nevertheless, we still have a conservation of the total spin pattern and could use the general form~\eqref{eq:slim_general_def} with $n=N_F$ to define conserved quantities accociated to this. 
Let us take $\hat{\mathcal{Q}}_{i_1\ldots i_{N_F}}$ to be the projector onto states where the $N_F$ fermions occupy the sites $i_1,\ldots,i_{N_F}$ and let $\mathbf{\sigma}$ be a cyclic permutation of the indices $i_1,i_2,\ldots,i_{N_F}$. Then the following choices all correspond to conserved quantities:
\begin{align*}
C^{(1)}_{i_1\ldots i_{N_F}} &= \sum_{\mathbf{\sigma}} S_{\mathbf{\sigma}(i_1)}^z, \\  
C^{(2)}_{i_1\ldots i_{N_F}} &= \sum_{\mathbf{\sigma}} S_{\mathbf{\sigma}(i_1)}^z S_{\mathbf{\sigma}(i_2)}^z, \\    
C^{(3)}_{i_1\ldots i_{N_F}} &= \sum_{\mathbf{\sigma}} S_{\mathbf{\sigma}(i_1)}^z S_{\mathbf{\sigma}(i_2)}^z S_{\mathbf{\sigma}(i_3)}^z, \\
&\vdots \\
C^{(N_F)}_{i_1\ldots i_{N_F}} &= \sum_{\mathbf{\sigma}} S_{\mathbf{\sigma}(i_1)}^z S_{\mathbf{\sigma}(i_2)}^z S_{\mathbf{\sigma}(i_3)}^z \ldots S_{\mathbf{\sigma}(i_{N_F})}^z.
\end{align*}
$C^{(1)}$ is just the total magnetization $S_{i_1}^{z} + \ldots + S_{i_{N_F}}^{z}$. $C^{(2)}$ measures the AFM ordering of the spins in squeezed space, etc. Note that they are not all independent, for example $C^{(N_F)}$, which measures the overall spin parity, is completely determined by $C^{(1)}$.

Nevertheless, while one can write conserved quantities for the periodic case, they are qualitatively very different from the SLIOMs of the open chain. The main difference is that in this case, with periodic boundaries, the conserved quantities do not factorize into products of 1-particle charges (SLIOMs with $n=1$). Instead, for a typical state, they involve a sum over all extensively many particles, and thus any notion of localization is lost.

\section{Averaging over ensembles of random states}\label{app:Haar}

Here we briefly summarize the relevant formulae for averaging over both Haar random states, as well as random states restricted to a fixed U(1) symmetry sector. 

\subsection{Haar average and variance}

A Haar random state $\ket{\psi}$ can be written as $\ket{\psi} = U\ket{0} = \sum_i U_{\alpha0}\ket{\alpha}$, where $U$ is a unitary matrix chosen from the Haar ensemble and $\ket{0}$ is an arbitrary basis element from a complete orthonormal basis $\{\ket{\alpha}\}$. The average of an an observable $\hat O$ is then  
\begin{equation} \label{eq:Aver}
\mathbb{E}_\text{Haar}[\braket{\psi|\hat O|\psi}] = \sum_{\alpha,\beta} O_{\alpha\beta} \mathbb{E}_\text{Haar}[U^*_{\alpha0} U_{\beta0}] = \frac{\text{tr}(\hat O)}{D},
\end{equation}
where $D$ is the Hilbert space dimension and we have used the fact that 
\begin{equation}
\mathbb{E}_\text{Haar}[U^*_{\alpha0} U_{\beta0}] = \frac{\delta_{\alpha\beta}}{D}.
\end{equation}

To get the variance over the Haar distribution, we are going to need to average over higher moments of the unitary $U$. In particular we have to evaluate
\begin{equation}
\mathbb{E}_\text{Haar}[\braket{\psi|\hat O|\psi}^2] = \sum_{\alpha\beta\mu\nu} O_{\alpha\beta} O_{\mu\nu} \mathbb{E}_\text{Haar}[U^*_{\alpha 0} U_{\beta 0} U^*_{\mu 0} U_{\nu 0}],
\end{equation}
which is given by the formula
\begin{equation}\label{eq:HaarU4}
\mathbb{E}_\text{Haar}[U^*_{\alpha 0} U_{\beta 0} U^*_{\mu 0} U_{\nu 0}] = \frac{ \delta_{\alpha\beta} \delta_{\mu\nu} + \delta_{\alpha\nu}\delta_{\beta\mu}}{D(D+1)}.
\end{equation}
Using this, one find that the variance is
\begin{multline}
\mathbb{E}_\text{Haar}[\braket{\psi|\hat O|\psi}^2] - \mathbb{E}_\text{Haar}[\braket{\psi|\hat O|\psi}]^2 = 
\\
= \frac{1}{D+1} \left[ \frac{\text{tr}(\hat O^2)}{D} - \left(\frac{\text{tr}(\hat O)}{D}\right)^2 \right]. 
\end{multline}
The particular cases we considered in the main text correspond to projection operators, $\hat O^2 = \hat O$. In this case, defining the probability $p = \text{tr}(\hat O) / D$ we get
\begin{equation}
\mathbb{E}_\text{Haar}[\braket{\psi|\hat O|\psi}^2] - \mathbb{E}_\text{Haar}[\braket{\psi|\hat O|\psi}]^2 = \frac{p - p^2}{D+1},
\end{equation}
which is suppressed by a factor of $D$ compared to $p$ itself.

\subsection{Fixed U(1) symmetry sectors}

In order to consider random states with a fixed eigenvalue under some U(1) symmetry, we should to consider a unitary $U$ that commutes with the symmetry operator. That is, we take $U$ to be block diagonal in the symmetry basis, with each block an independent Haar random unitary. In this case, we can average over the block separately. Denoting the U(1) quantum numbers by $N$, we then get a generalization of the previous formula,
\begin{equation}
\mathbb{E}_\text{U(1)}[U^*_{\alpha\alpha'} U_{\beta\beta'}] = \sum_N \frac{P^{(N)}_{\alpha\beta} P^{(N)}_{\alpha'\beta'}}{D_N},
\end{equation}
where $P^{(N)}$ is a projector onto the symmetry sector with $N$, and $D_N \equiv \text{tr}(P^{(N)})$ is the corresponding dimension. 

The ensemble of random states is defined by $\ket{\psi} = U \ket{0}$ where the basis state $\ket{0}$ is  chosen to have a fixed quantum number $N$. This picks out a single projector from the above sum to give
\begin{equation}
\mathbb{E}_\text{U(1)}[\braket{\psi|\hat O|\psi}] = \frac{\text{tr}(\hat O P^{(N)})}{D_N}.
\end{equation}
In the cases we consider, $\hat O$ and $P^{(N)}$ are both diagonal projectors in the same local product basis. $\text{tr}(\hat O P^{(N)})$ is therefore simply given by counting the number of configurations that are in the intersection, satisfying both $\hat O = 1$ and $P^{(N)} = 1$. 

In an analogous manner, one could calculate variances over this ensemble. The result is the same as for the Haar random case, but with $D \to D_N$ and the $\delta$ functions in Eq.~\eqref{eq:HaarU4} replaced by matrix elements of $P^{(N)}$. Consequently, the variance becomes
\begin{multline}
\mathbb{E}_\text{U(1)}[\braket{\psi|\hat O|\psi}^2] - \mathbb{E}_\text{Haar}[\braket{\psi|\hat O|\psi}]^2 = 
\\
= \frac{1}{D_N+1} \left[ \frac{\text{tr}(\hat O P^{(N)} \hat O P^{(N)})}{D_N} - \left(\frac{\text{tr}(\hat O P^{(N)})}{D_N}\right)^2 \right]. 
\end{multline}
As mentioned above, we are interested in cases where $\hat O$ and $P^{(N)}$ are both projectors, diagonal in the same basis. Therefore $\hat O P^{(N)}$ is also a projector, $\hat O P^{(N)} \hat O P^{(N)} = \hat O P^{(N)}$, and the variance is now suppressed by a factor of $D_N + 1$, still exponentially large in system size for typical symmetry sectors.

\subsection{Computation of the average number of defects} \label{app:applic}

As an application, in this section we compute the average filling fraction of defects $\avg{\nu_d}$. To do so, let us compute $\mathbb{E}_\text{Haar}[\avg{\psi|N^d|\psi}]$ appearing in Eq.~\eqref{eq:Nd}. Using Eq.~\eqref{eq:Aver} we find that
\begin{align} 
 &\mathbb{E}_\text{Haar}[\avg{\psi|N^d|\psi}]=\frac{1}{2}\sum_{k=1}^L  \frac{\text{tr}\Big[\big(\hat q_{k+1}\big)^2\big(1+\hat q_k \hat q_{k+1} \big)\Big]}{3^L}.
\end{align}

Now we split the computation in two steps. First, let us compute each of the terms individually:
\begin{align} \nonumber
&\frac{1}{3^L}\text{tr}\Big[\big(\hat q_{k+1}\big)^2\Big]=\frac{1}{3^L}\sum_{i,j}\text{tr}\Big[\hat{\mathcal{P}}^{k+1}_i\hat{\mathcal{P}}^{k+1}_j S_i^z S_j^z \Big]\\
&=\frac{1}{3^L}\sum_{i}\text{tr}\Big[\hat{\mathcal{P}}^{k+1}_i \big(S_i^z\big)^2 \Big]=\sum_ip_\text{Haar}(i;k+1),
\end{align}
where we have used that for $i < j$ the trace vanishes due to $\text{tr}(S_j^z) = 0$ and $p_\text{Haar}$ is defined in Eq.~\eqref{eq:Haaravg}. Now, combining this with the fact tha $\hat{\mathcal{P}}^{k}_i \hat{\mathcal{P}}^{k+1}_i=0$, the second term vanishes:
\begin{align} \nonumber
&\frac{1}{3^L}\text{tr}\Big[\hat q_k \big(\hat q_{k+1}\big)^3 \Big]=\frac{1}{3^L}\sum_{i,j}\text{tr}\Big[\hat{\mathcal{P}}^{k}_i \hat{\mathcal{P}}^{k+1}_j S_i^z S_j^z \Big]\\
&=\frac{1}{3^L}\sum_{i}\text{tr}\Big[\hat{\mathcal{P}}^{k}_i \hat{\mathcal{P}}^{k+1}_i \big(S_i^z\big)^2 \Big]=0.
\end{align}

Thus, for a Haar random state with filling fraction $\nu=2/3$, and using the fact that $\sum_{k}p_\text{Haar}(i;k+1)=\nu$, we obtain that the typical filling fraction of defects is
\begin{align} \nonumber
&\avg{\nu_d}=\frac{1}{L}\mathbb{E}_\text{Haar}[\avg{\psi|N^d|\psi}]=\frac{1}{2L}\sum_{k=1}^L\sum_ip_\text{Haar}(i;k+1)\\ 
&=\frac{1}{L}\sum_i \frac{\nu}{2} =\frac{1}{3}.
\end{align}
Intuitively, this comes from the fact that any given charge has equal probability of having the same vs. opposite sign as the nearest charge on the left, making the probability of finding a defect on a particular site $\nu / 2 = 1/3$.

\section{Evaluating SLIOMs in eigenstates}\label{app:eigstates}

When discussing the spatial distribution of SLIOMs in the main text, we used ensembles of random states (either with or without fixing the total number of particles). As we showed, the variance over different choices of random states is exponentially small in system size, implying that averaging over the ensemble indeed provides an extremely good approximation of the expectation value of ${\mathcal{O}^k_i}^\dagger \mathcal{O}^k_i$ for most states within the same Hilbert space. Nevertheless, one might wonder what these distributions look like for specific eigenstates of the Hamiltonian $H_{t-J_z}$. Here we address this question.

\begin{figure}[t]
	\includegraphics[width=1.0\columnwidth]{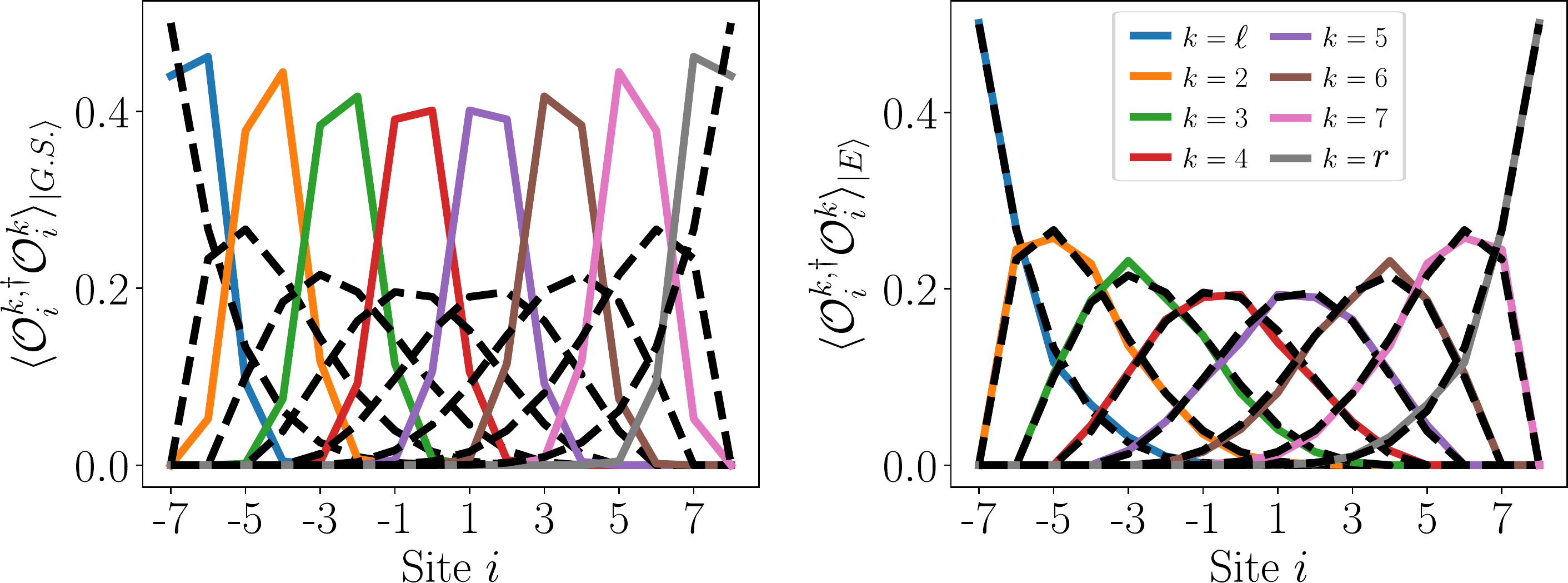}
	\caption{\textbf{Spatial distribution of SLIOMs for energy eigenstates.} Spatial distribution of the expectation value $\braket{\psi|{\mathcal{O}^k_i}^\dagger \mathcal{O}^k_i|\psi}$, for the ground state (left) and an excited state (right) within the sector $N_F = L/2$, $S^z_\text{tot}=0$. The excited state is randomly picked from within a sector with a fixed spin pattern (see main text). Both states correspond to (partially) localized distributions. For the excited state, this is close to the Haar average (dashed lines), while for the ground state the distribution is more tightly localized. }
	\label{fig:eigenstates}
\end{figure}

In particular, we fix a global symmetry sector with half filling ($N_F = L/2$) and total magnetization $S^z_\text{tot} = 0$. We consider two eigenstates within this sector: i) the ground state, that has the lowest energy within this symmetry sector and ii) a randomly chosen, highly excited eigenstate within the fixed spin pattern sector corresponding to $\gamma_k=1$ (spins pointing up) for $k\leq N_F/2$ and $\gamma_k=-1$ (spins pointing down) for $k>N_F/2$. In the latter case, we expect to be close to the typical (Haar random) state with the same $N_F$, which we considered in the main text. Indeed, as shown in Fig.~\ref{fig:eigenstates}(b), we find that the distribution of $\braket{{\mathcal{O}^k_i}^\dagger \mathcal{O}^k_i}$ is well approximated by Eq.~\eqref{eq:U1Haaravg}, up to finite size corrections. The ground state, on the other hand, is a highly atypical state (for example it does not have a volume law entanglement). For this reason, the distribution is noticeably different from the Haar average. Nevertheless, we find that it is in fact \emph{more} tightly localized, as one can observe from Fig.~\ref{fig:eigenstates}(a). Thus the statistical localization of the conserved quantities remains valid also when considered this state.

\section{Spatially resolved autocorrelations at long times}

As noted in the main text (see Eq.~\eqref{eq:mazur}), Mazur's inequality provides a strict lower bound on autocorrelations $\braket{S_j^z(t)S_j^z}$ in terms of the conserved quantities of the system. However, to understand the spatial spreading of spin density $S_j^z$, it is also interesting to consider correlations between different sites of the form $\braket{S_j^z(t)S_i^z}$, for which the same lower bound does not exist. Here we provide a conjecture for the long-time average of these correlations in the thermodynamic limit of the $t-J_z$ model and show some supporting numerics.

While one cannot lower bound the correlations between different sites in the same way as autocorrelators, one could in principle calculate their time average if one had access to a \emph{complete} orthogonal set of $3^L$ conserved quantities (a basis of all operators diagonal in the eigenbasis of $H_{t-J_z}$). Given such an orthogonal set $\{\hat{I}_a\}_{a=1}^{3^L}$, one can prove~\cite{Suzuki71} that the time average becomes
\begin{multline}\label{eq:mazur_offdiag}
\lim_{T\to\infty} \frac{1}{T} \int\text{d}t \, \braket{S_j^z(t)S_i^z}_{\beta=0} = \sum_a \frac{\braket{S_j^z\hat{I}_a}_{\beta=0} \braket{S_i^z \hat{I}_a}_{\beta=0}}{\braket{\hat{I}_a^2}_{\beta=0}}
\end{multline}

\begin{figure}[t]
	\includegraphics[width=0.85\columnwidth]{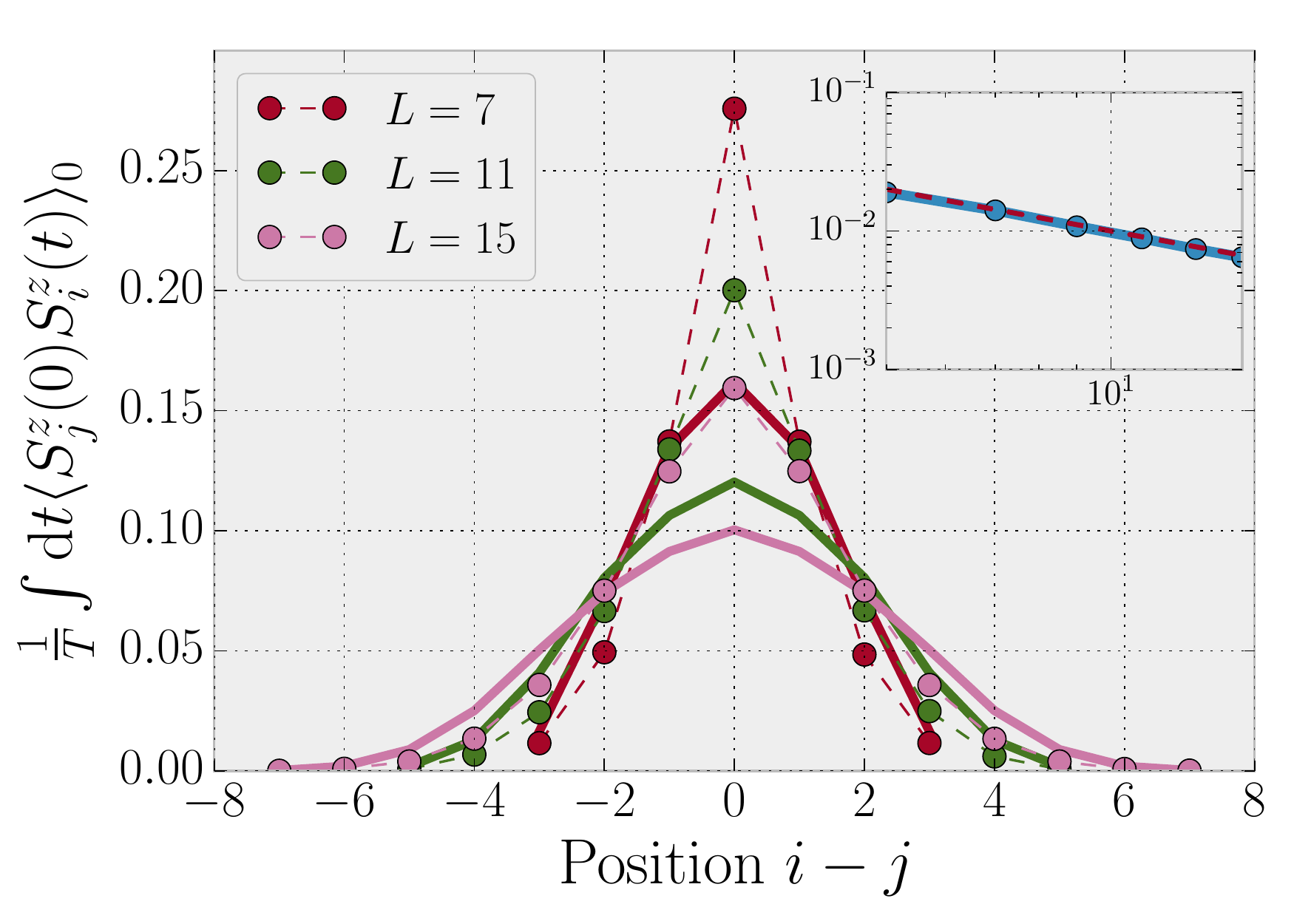}
	\caption{\textbf{Time averaged correlations vs. their conjectured values.} The dots (connected by narrow dashed lines) show the long-time average (averaged between times $t=50$ and $t=100$) of the correlator $\braket{S_j^z(t)S_i^z}_{\beta=0}$, while the solid lines represent $C_{ij}(\infty)$, defined by the formula~\eqref{eq:mazur_offdiag}. This is a lower bound near the origin, but becomes smaller then the numerical value in the tails (i.e., the observed distribution is actually \emph{narrower} than the prediction). However, the two curves approach each other as system size is increased. This is shown by the inset, where the blue dotted curve represents $\sum_i \left[ \frac{1}{T} \int_{50}^{100} \text{d}t \braket{S_j^z(t)S_i^z}_{\beta=0} - C_{ij}(\infty) \right]^2$ as a function of $L$, approximately decreasing as $1/L$ (red dashed line).}
	\label{fig:mazur_spatial_App}
\end{figure}

\begin{figure}[h!]
	\includegraphics[width=0.49\columnwidth]{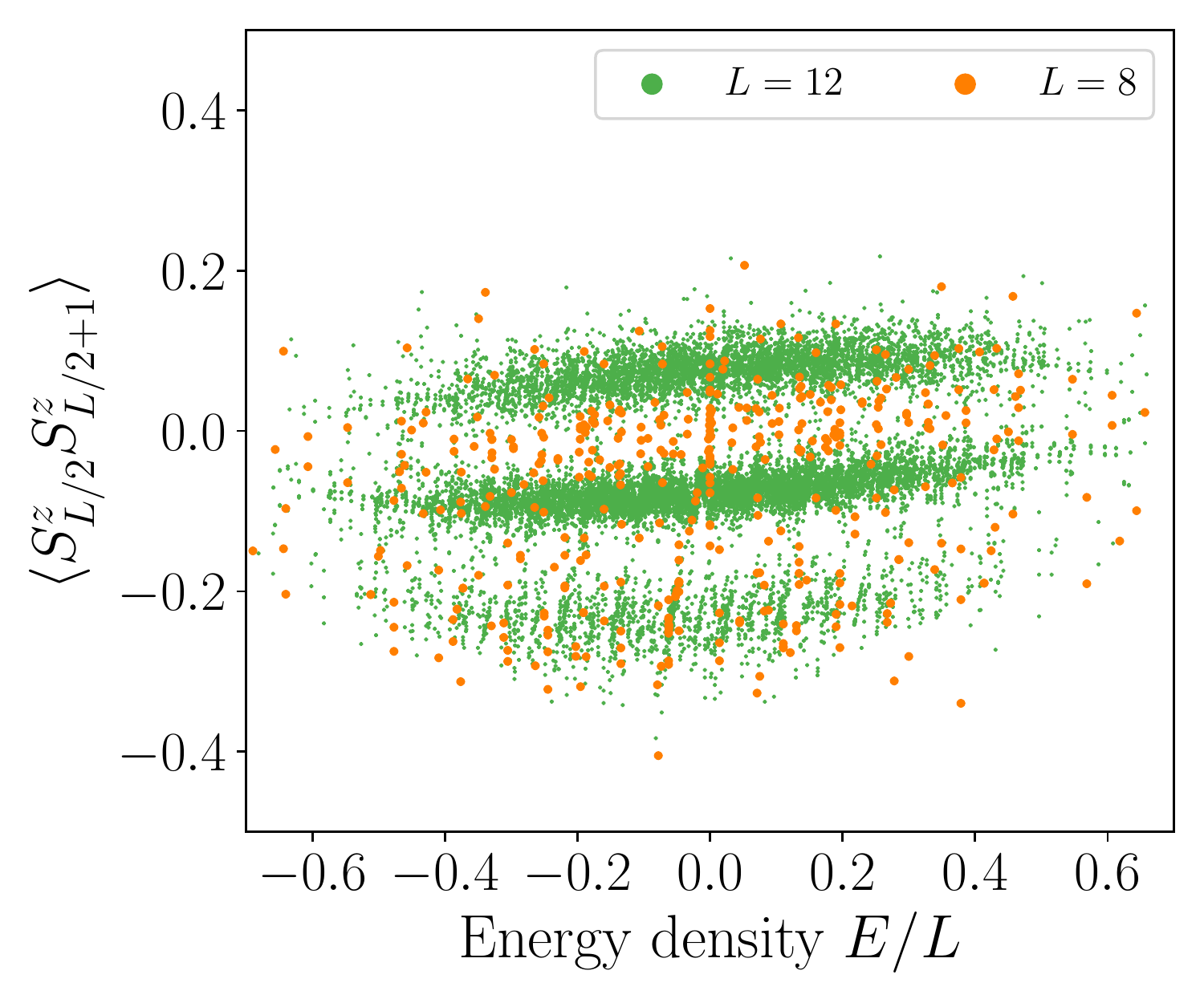}
	\includegraphics[width=0.49\columnwidth]{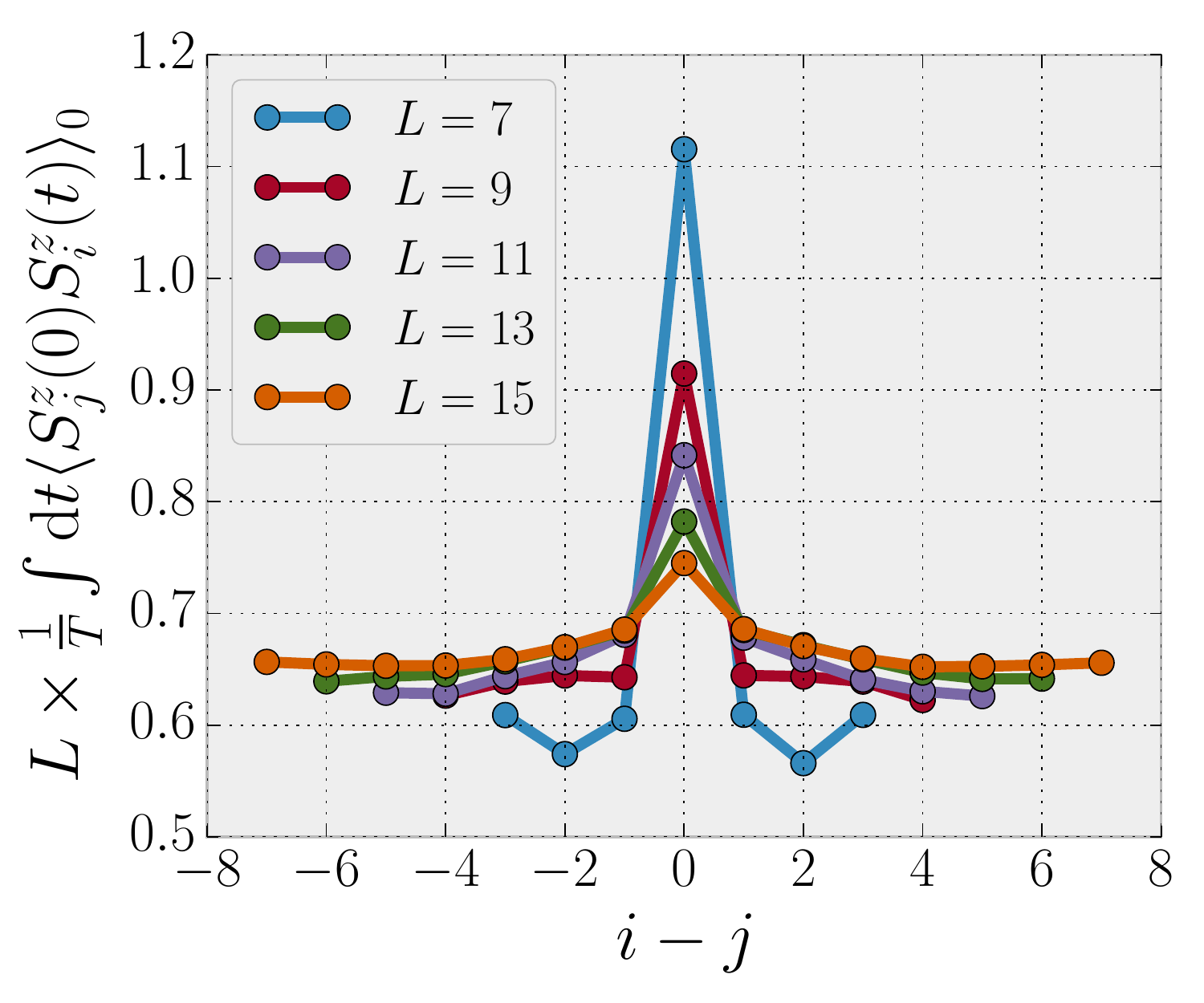}
	\caption{\textbf{Thermalization in the $t-J_z$ model with closed boundaries} Left: expectation values of nearest neighbor antiferromagnetic correlations in eigenstates for $N_F = L/2$, $S^z_\text{tot}=0$ for different system sizes. The distribution has a width that does not decrease with system size. Right: time average (between times $50$ and $100$) of the spatially resolved spin-spin correlations at infinite temperature. While there is a small peak around the origin remaining for the available system sizes, the correlations mostly spread out over the whole chain and take values $\propto 1/L$, unlike the case of an open chain shown in Fig.~\ref{fig:tJz_autocorr}(c).}
	\label{fig:tjZ_closed}
\end{figure}

The formula~\eqref{eq:mazur_offdiag} requires knowledge of exponentially many conserved quantities, which is much more than the information contained in only the SLIOMs $\hat{q}_k$ defined in Eq.~\eqref{eq:qk_def}. Our conjecture is that in the limit $L\to\infty$ the correct time average is given by restricting the sum on the right hand side to the set $\{\hat{q}_k\}$, ignoring other conserved quantities, i.e.
\begin{equation}
\sum_k \frac{\braket{S_j^z\hat{q}_k}_{\beta=0} \braket{S_i^z \hat{q}_k}_{\beta=0}}{\braket{\hat{q}_k^2}_{\beta=0}} \equiv C_{ij}(\infty).
\end{equation}
Indeed, this conjecture is supported by the observation that the quantities $C_{ij}(\infty)$ are all positive and they sum up to the correct value, $\sum_j C_{ij}(\infty) = 2/3 = \sum_j \braket{S_j^z(t)S_i^z}_{\beta=0}$. 
This means that the contribution coming from all remaining terms ($\hat{I}_a \neq \hat{q}_k$) in the sum~\eqref{eq:mazur_offdiag} have to be such that their sum over $i$ vanishes. Our conjecture amounts to saying that they in fact all individually vanish in the thermodynamic limit.

This conjecture is supported by our small scale numerics, which show that the difference between the two distributions decreases with $L$. In particular, we can define the mean square distance of the two,
\begin{equation}
\sum_i\left[\frac{1}{T} \int \text{d}t \braket{S_j^z(t)S_i^z}_{\beta=0} - C_{ij}(\infty)\right]^2.
\end{equation}
We find (see in particular the inset of Fig.~\ref{fig:mazur_spatial_App}) that this quantity decreases with system size, approximately as $1/L$.
Note that the distribution $C_{ij}(\infty)$ has a width $\propto \sqrt{L}$, such that our conjecture implies that for a finite open chain the charge remains trapped in a region much smaller than the entire system as discussed also in the main text. 

\section{$t-J_z$ model with closed boundaries}\label{app:pbc}

Our discussion of the $t-J_z$ model in the main text focused on a chain with open boundaries. This allowed us to label fermions by an integer $k$, starting from one of the endpoints, leading to the definition of SLIOMs in Eq.~\eqref{eq:qk_def}. Here we detail how the situtation changes when periodic boundary conditions are taken. 

In the periodic case, the conserved spin pattern is only well defined modulo cyclic translations around the chain, allowing for additional matrix elements between certain sectors that are disconnected for the open chain. Nevertheless, this only reduces the number of disconnected sectors by at most a factor of $1/L$, such that there are still exponentially many invariant subspaces and the dimension of the largest one still scales asymptotically as $\sim 2^L$. The Hilbert space is therefore still strongly fragmented and should therefore violate ETH. Indeed, repeating the same calculation as in Fig.~\ref{fig:eth_open}(b) for the closed chain, we again find a wide distribution of diagonal matrix elements of $S^z_{L/2}S^z_{L/2=1}$. This is shown in Fig.~\ref{fig:tjZ_closed}(a). Approximating eigenstates by an equal weight superposition of hole positions in this case suggests that the width of the distribution asymptotically decreases with system size as $L^{-1/2}$ in the thermodynamic limit (recall, that for open chains the narrowing was slower, $\sim L^{-1/4}$). 

The difference between open and closed boundaries becomes even more explicit when we consider the conserved quantities that label the disconnected sectors. In particular, the SLIOMs defined in Eq.~\eqref{eq:qk_def} are no longer conserved, since fermions can now circle around the boundaries. Indeed, while the whole of the spin pattern is still conserved, talking about the spin of individual fermions is no longer meaningful and consequently, the spatial localization associated to the conserved quantities breaks down. This explains the different asymptotic scaling in the width of the distribution of diagonal matrix elements. It also shows up when considering the late-time behavior of correlations of the form $\braket{S_j^z(t)S_i^z}_{\beta=0}$. Unlike the case with open boundaries, where these spread out only over a region of size $\sqrt{L}$ (see Fig.~\ref{fig:tJz_autocorr}(c)), for a closed chain the spread out over the entire chain, saturation to a value of $O(1/L)$. This is shown in Fig.~\ref{fig:tjZ_closed}(b).

Note that for the Hamiltonian $H_3$ in Eq.~\eqref{eq:H3} the situation is quite different. While labeling individual defects also loses meaning with periodic boundaries, the regions surrounded by neighboring defects are still well defined and have the same $O(1)$ size as with open boundaries. This is consistent with the localized behavior (i.e., infinitely long-lived autocorrelations) \emph{in the bulk}, discussed in Sec.~\ref{sec:local_dipoles}.

\section{Saturation value of the Entanglement entropy} \label{app:ScaEE}
\begin{figure}
	\centering
	\includegraphics[width=0.7\linewidth]{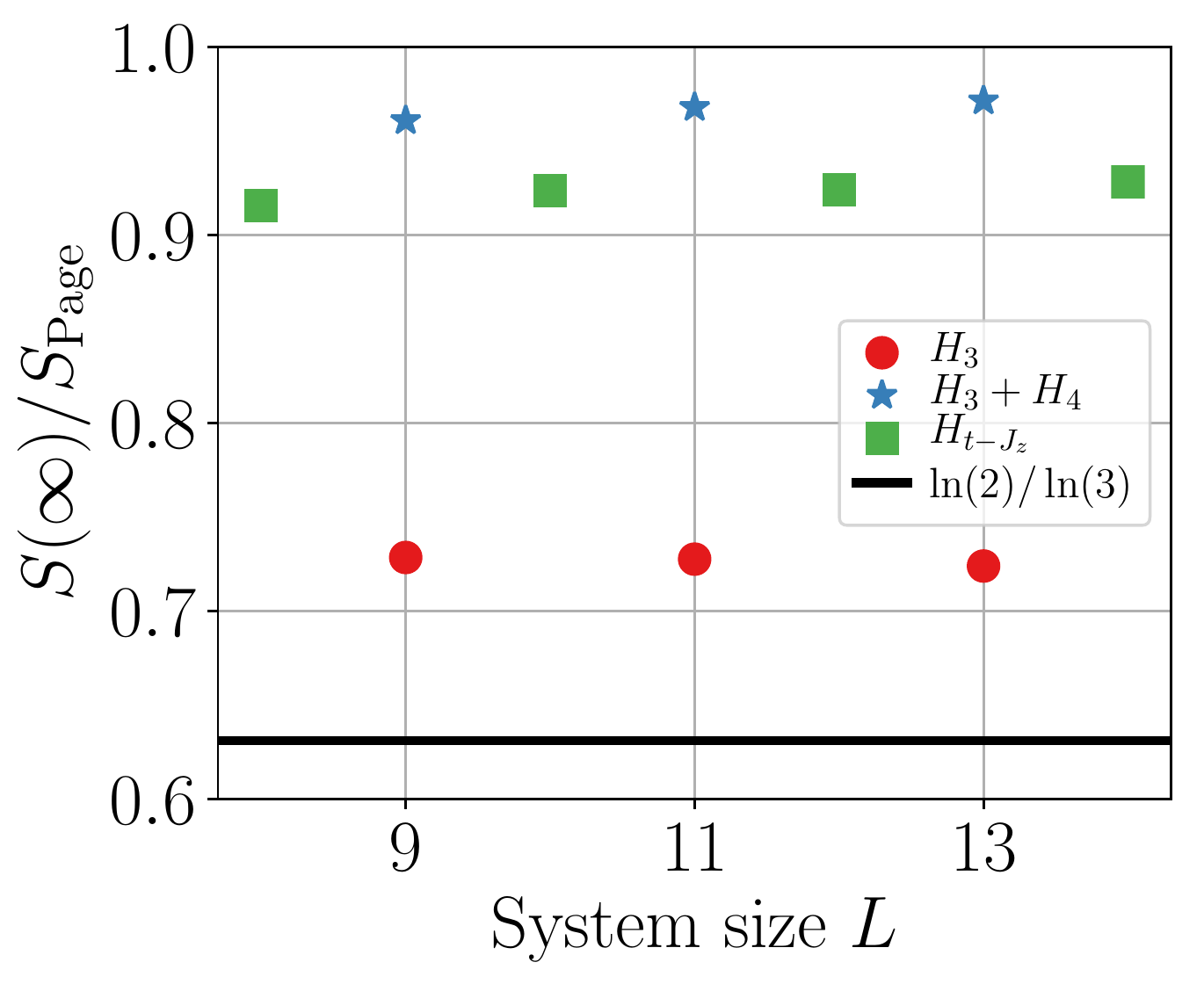}
	\caption{\textbf{Scaling of the saturation value of the entanglement entropy with system size.} We show the ratio of the saturation value of the entanglement entropy $S(\infty)$ and the Page value $S_{\text{Page}}=\ln(3)L/2 -1/2$, for for $H_3$ (red circles), $H_3+H_4$ (blue stars) and $H_{t-J_z}$ (green squares). }
	\label{fig:scalingee}
\end{figure}

In this appendix we provide the data obtained for the scaling of the saturation value of the entanglement entropy ($S(\infty)$) with initial (Haar) random product states (not in the $z$ basis), for the models studied in the main text. The data for the Hamiltonian $H_3$, was provided in App.~C of Ref.~\onlinecite{Sala19}, while a random unitary circuit model with the same symmetries was studied in Ref.~\onlinecite{Vedika19}. For completeness, we also show the scaling for the dipole-conserving Hamiltonian $H_3+H_4$ with 
\begin{equation}
H_4 = -\sum_n \Big[S_n^+S_{n+1}^-S_{n+2}^-S_{n+3}^+ + \text{H.c.} \Big],
\end{equation}
which is only weakly fragmented and saturates close to the Page value~\cite{Page93}, $S_{\text{Page}}=\ln(3)L/2 -1/2$, up to a constant offset.

In Fig.~\ref{fig:scalingee}, we show the scaling of $S(\infty)$ with system size for $H_3$ (red circles), $H_3+H_4$ (blue stars) and $H_{t-J_z}$ (green squares).
The scaling (for the small system sizes the simulations were performed) suggests that for the $t-J_z$ model $S(\infty)$ will approach $S_{\text{Page}}$ in the thermodynamic limit, while it remains only a fraction of it for $H_3$.

\bibliography{slim_0220.bbl}

\end{document}